\documentclass{article}
\usepackage[english]{babel}
\usepackage[letterpaper,top=1.8cm,bottom=1.8cm,left=2.6cm,right=2.6cm,marginparwidth=1.75cm]{geometry}
\usepackage{amsmath,amsfonts}
\usepackage{graphicx}
\usepackage{array}
\usepackage{amssymb}
\usepackage{longtable}
\usepackage{caption}
\usepackage{amsmath}
\usepackage{float}
\usepackage[colorlinks=true, allcolors=blue]{hyperref}
\usepackage{natbib}

\usepackage{booktabs}
\usepackage{tabularx}
\usepackage{subfigure}
\usepackage{multirow}
\usepackage{makecell}
\usepackage{algorithm}
\usepackage{algpseudocode}

\DeclareCaptionLabelFormat{cont}{#1~#2\alph{ContinuedFloat}}
\setcitestyle{authoryear,round}
\providecommand{\keywords}[1]
{\small	\textbf{Keywords:} Speed limit control, connected and automated vehicles, freeway network, traffic composition uncertainty, freeway congestion}
\date{}
\title{Dynamic Speed Limit Control of Connected Automated Vehicles in Freeway Networks Considering Traffic Composition Uncertainty}
\begin{document}
\maketitle
\vspace{-4em}
\begin{center}
Lei Wei$^{1,2}$, Yu Han$^{3}$, 
Haiyang Yu$^{*}$$^{1,2}$, Yunpeng Wang$^{*}$$^{1,2}$
\end{center}
\footnotetext[1]{State Key Laboratory of Intelligent Transportation System, Beihang University, 100191, Beijing, China.}
\footnotetext[2]{School of Transportation Science and Engineering, Beihang University, 100191, Beijing, China.}
\footnotetext[3]{Thrust of Intelligent Transportation, The Hong Kong University of Science and Technology (Guangzhou), 511453, Guangzhou, China.}

\renewcommand{\thefootnote}{\fnsymbol{footnote}}
\footnotetext[1]{Corresponding author: hyyu@buaa.edu.cn; ypwang@buaa.edu.cn}
\renewcommand{\thefootnote}{\arabic{footnote}}

\begin{abstract}
Dynamic speed limit control has emerged as a promising strategy to improve freeway sustainability in mixed traffic environments with connected automated vehicles (CAVs). However, most existing approaches assume that the CAV penetration rate is deterministic and can be accurately known throughout the control horizon. In reality, the penetration rate has  inherent observation errors, leading to uncertainty in mixed traffic composition, which in turn degrades control performance. To overcome this limitation, this study proposes a novel model predictive control (MPC)  framework for dynamic CAV speed limit control in freeway networks that explicitly incorporates traffic composition uncertainty into both flow prediction and control optimization. An uncertainty-aware macroscopic mixed traffic model is first developed, where the uncertain penetration rate propagates through the mixed fundamental diagram to the flow dynamics by affecting the mixed free-flow speed, capacity, and capacity drop condition. Then, a traffic composition-aware MPC is formulated to optimize CAV speed limits against multiple admissible penetration rate realizations, thereby improving control robustness under heterogeneous traffic conditions. Simulation experiments are conducted on both a single-bottleneck freeway corridor and a multi-bottleneck freeway network with merge-diverge interactions. The results demonstrate that the proposed controller generates more spatially coordinated speed limits, which effectively reduce travel time spent and provide environmental benefits. 
\end{abstract}
\section{Introduction}
Freeway congestion remains one of the major challenges in modern transportation systems, resulting in increased travel time, fuel consumption, and traffic emissions \citep{wang2020trajectory,wei2022turn}. With the rapid deployment of connected automated vehicles (CAVs), dynamic speed control has emerged as a promising traffic management strategy because CAVs can accurately follow infrastructure-issued speed commands through vehicle-to-infrastructure (V2I) communication \citep{tu2019quantifying,wei2024hierarchical,niroumand2020joint,ma2021eco}. Compared with conventional variable speed limit (VSL) control designed for human-driven traffic \citep{khondaker2015variable,lu2023variable,wei2025first}, CAV speed control enables more coordinated traffic regulation and provides new opportunities for mitigating congestion propagation, improving bottleneck throughput, and enhancing overall freeway efficiency \citep{li2023developing}. Consequently, model-based dynamic speed control has attracted increasing attention in recent years.


\subsection{Related work}
Model predictive control (MPC) has become one of the most widely adopted approaches for freeway speed limit control because it explicitly predicts future traffic evolution and optimizes control actions over a finite prediction horizon \citep{luo2025distributed,pham2023distributed,liberati2024single}. Depending on the traffic flow model adopted for state prediction, existing MPC approaches can generally be classified into microscopic and macroscopic methods. Microscopic models, such as car-following models, explicitly describe individual vehicle behaviors and can accurately reproduce vehicle interactions \citep{khondaker2015variable}. However, their computational complexity increases rapidly with network size, making them unsuitable for real-time network-level control. Consequently, most existing studies employ macroscopic traffic flow models to balance prediction accuracy and computational efficiency \citep{jin2025variable,mao2022variable,han2017resolving,li2020research}.

Among macroscopic models, the second-order METANET model has been widely used for freeway speed control. Several studies developed nonlinear MPC frameworks based on METANET to optimize VSLs for reducing travel time and alleviating freeway congestion \citep{wang2022macroscopic,frejo2014hybrid}. Although these approaches have demonstrated satisfactory control performance, the resulting optimization problem is nonlinear and computationally demanding, particularly for large-scale freeway networks \citep{imran2025mpc}. Another widely adopted class of prediction models is the cell transmission model (CTM). Owing to its piecewise linear formulation, CTM-based MPC significantly improves computational efficiency and has been successfully applied to freeway congestion mitigation, capacity drop prevention, and coordinated VSL control \citep{mao2022variable,jin2025variable}. More recently, several studies further extended CTM-based control to mixed traffic environments with CAVs \citep{farda2026multiclass,zhang2025impacts,di2023integrated}, where dynamic speed control can be used to improve traffic efficiency and safety simultaneously. Nevertheless, CTM requires spatial discretization of every road segment, resulting in a large number of state variables and accumulated discretization errors for large freeway networks.

To improve computational efficiency while preserving traffic flow accuracy, link-based flow models have recently attracted increasing attention. Compared with CTM, the link transmission model (LTM) reproduces kinematic-wave propagation without dividing roads into multiple cells, thereby substantially reducing the model dimension \citep{jin2015continuous,lu2024link}. Several studies have incorporated LTM into MPC frameworks for freeway speed control and demonstrated its computational advantages. \cite{wei2025first} extended the standard LTM to explicitly model the propagation of VSL information along freeway links, enabling more realistic representation of speed-control effects. \cite{hajiahmadi2015integrated} integrated the LTM into an MPC framework for coordinated freeway control with VSLs and ramp metering, demonstrating improved computational efficiency compared with cell-based formulations. Owing to the reduced number of state variables and larger admissible simulation time steps, LTM-based MPC provides a promising framework for network-level freeway traffic control. However, these studies generally assume deterministic traffic conditions and optimize speed limits based on perfectly known flow parameters.

Compared with conventional VSL control designed for human-driven traffic, the rapid deployment of CAVs has motivated researchers to extend freeway traffic modeling and control frameworks to mixed flow environments, where CAVs and human-driven vehicles (HVs) coexist. Existing studies have demonstrated that the CAV penetration rate significantly affects macroscopic traffic characteristics, including equilibrium headway, free-flow speed, capacity, traffic stability, and congestion formation \citep{jin2026optimal,gong2024modeling,zhang2024data,qin2025management}. Accordingly, various mixed traffic fundamental diagrams have been proposed to describe the relationship between flow variables under different CAV penetration rates \citep{zhang2025stochastic,bilal2023evaluation}. These models provide an effective foundation for model-based traffic prediction and control in mixed  environments. At the macroscopic level, \cite{zhou2020modeling} developed one of the earliest mixed fundamental diagrams by explicitly incorporating stochastic headways and CAV platooning into the flow–density relationship, demonstrating that increasing CAV penetration not only improves capacity but also reduces the dispersion of the fundamental diagram. Building upon this work, \cite{yao2022fundamental} further extended the mixed fundamental diagram by considering platoon size and platoon intensity, and revealed the trade-off between capacity and stability under different CAV penetration levels. More recently, several studies have developed macroscopic mixed traffic models in which capacity, free-flow speed, or equilibrium headway vary with the CAV penetration rate, providing an effective basis for flow prediction and model-based control in heterogeneous traffic environments \citep{ma2023mixed,zhang2024discrete}.

Despite these advances, most existing mixed traffic flow models assume that the CAV penetration rate is accurately known and remains deterministic throughout the prediction horizon. In practical freeway operation, the penetration rate continuously varies over both space and time and can only be estimated from incomplete observations collected through connected vehicle messages or roadside sensing systems. Consequently, uncertainty in penetration estimation directly propagates to the mixed flow characteristics, resulting in uncertain flow dynamics. To improve controller performance under uncertain traffic conditions, robust traffic control has been extensively investigated \citep{seilabi2023robust,tan2025connected}. \cite{liu2022scenario} proposed a  distributed MPC framework for large-scale freeway control by representing uncertain demand and weather conditions with multiple scenarios, thereby improving the robustness of coordinated ramp metering and VSL control. Hierarchical robust MPC has also been developed to mitigate the influence of uncertain demand on freeway operation while maintaining computational efficiency through multi-level optimization with different objectives \citep{wei2025hierarchical}. In addition, stochastic MPC has been employed to explicitly account for probabilistic traffic disturbances and measurement uncertainty, enabling a trade-off between control performance and robustness under uncertain traffic flow evolution \citep{pham2023distributed}.

However, uncertainty in traffic composition fundamentally differs from the uncertainty considered in existing robust freeway control studies. Rather than acting as an additive disturbance, traffic composition uncertainty directly impacts the underlying mixed flow characteristics by affecting the free-flow speed, capacity, and congestion formation. Consequently, uncertainty in traffic composition changes the flow model itself rather than merely perturbing its states. Existing robust MPC frameworks generally do not explicitly capture this uncertainty propagation mechanism, leading to degraded prediction accuracy and control performance when the actual CAV penetration rate deviates from its nominal estimate.

Recently, artificial intelligence (AI)-based approaches have attracted increasing attention in freeway traffic control. Deep reinforcement learning (DRL) has been widely applied to VSL control, ramp metering, and integrated freeway traffic management owing to its capability of learning complex control policies directly from traffic data without requiring explicit traffic flow models \citep{han2026cooperative,han2022new,jin2025variable}. Despite these advances, existing AI-based approaches generally require large amounts of representative training data \citep{elsamadisy2025deep}. They often suffer from limited interpretability and generalization when traffic conditions differ from those encountered during training. More importantly, adapting learned control policies to time-varying mixed traffic composition remains challenging.

In summary, although considerable progress has been made in both model-based and data-driven freeway speed control methods, existing studies primarily address uncertainties arising from traffic demand, model mismatch, measurement errors, or external disturbances. In contrast, uncertainty originating from the mixed traffic flow composition, particularly the time-varying and imperfectly observed CAV penetration rate, has received little attention. As a result, a physically interpretable and computationally efficient control framework that explicitly incorporates penetration rate uncertainty into both flow prediction and control optimization is still lacking.


\subsection{Contributions}

This study proposes a novel MPC framework for dynamic CAV speed limit control in freeway networks considering traffic composition uncertainty. The main contributions are summarized as follows:




Firstly, existing dynamic speed control methods typically assume deterministic mixed traffic characteristics and neglect the impact of uncertain CAV penetration rates on traffic flow dynamics. To address this limitation, an uncertainty-aware macroscopic mixed flow model is developed. In the proposed model, the CAV penetration rate is treated as an uncertain and time-varying traffic composition variable rather than a fixed parameter. Its uncertainty changes the mixed free-flow speed, mixed capacity, and capacity drop condition, thereby influencing the evolution of freeway traffic states.



Secondly, unlike previous studies that model uncertainty as exogenous disturbances or additive process noise, this study introduces a novel physically interpretable uncertainty propagation framework in which CAV penetration uncertainty propagates through the mixed fundamental diagram to traffic flow evolution. This establishes a direct connection between uncertain traffic composition and macroscopic flow dynamics, enabling more reliable prediction of mixed traffic dynamics.

Thirdly, a traffic composition-aware MPC framework is proposed that explicitly incorporates penetration uncertainty into the prediction and optimization process, enabling dynamic CAV speed control that mitigates congestion propagation and capacity drop while maintaining robust performance in heterogeneous mixed traffic environments.

The rest of this paper is organized as follows. Section \ref{framework label} presents the overall framework of the proposed dynamic speed limit control strategy. Section \ref{LTM_Formulation} introduces the mixed traffic flow model, including the mixed traffic fundamental diagram, and the capacity drop formulation. Section \ref{Controller formulation} and \ref{Solution method} formulates the MPC controller for handling traffic composition uncertainty. Section \ref{simulation} presents the simulation studies, where the proposed approach is compared with benchmark speed control strategies to demonstrate its effectiveness. Finally, Section \ref{Conclusion} concludes the paper and discusses future research directions.

\section{Formulation of the dynamic speed limit control for CAVs in freeway networks}
This section first provides an overview of the proposed MPC framework for freeway networks with mixed CAV and HV traffic flow. It then presents the link-based mixed flow model used for prediction, including the inflow and outflow formulations and the capacity drop mechanism, which capture the effects of CAV–HV interactions on traffic propagation while explicitly accounting for uncertainty in the CAV penetration rate. Finally, the design and solution procedure of the MPC controller are described for optimizing dynamic CAV speed limits under traffic composition uncertainty.

\subsection{Framework} 
\label{framework label}
\begin{figure}[H]
\captionsetup{font={small}}
\centering
\includegraphics[width=6.4in]{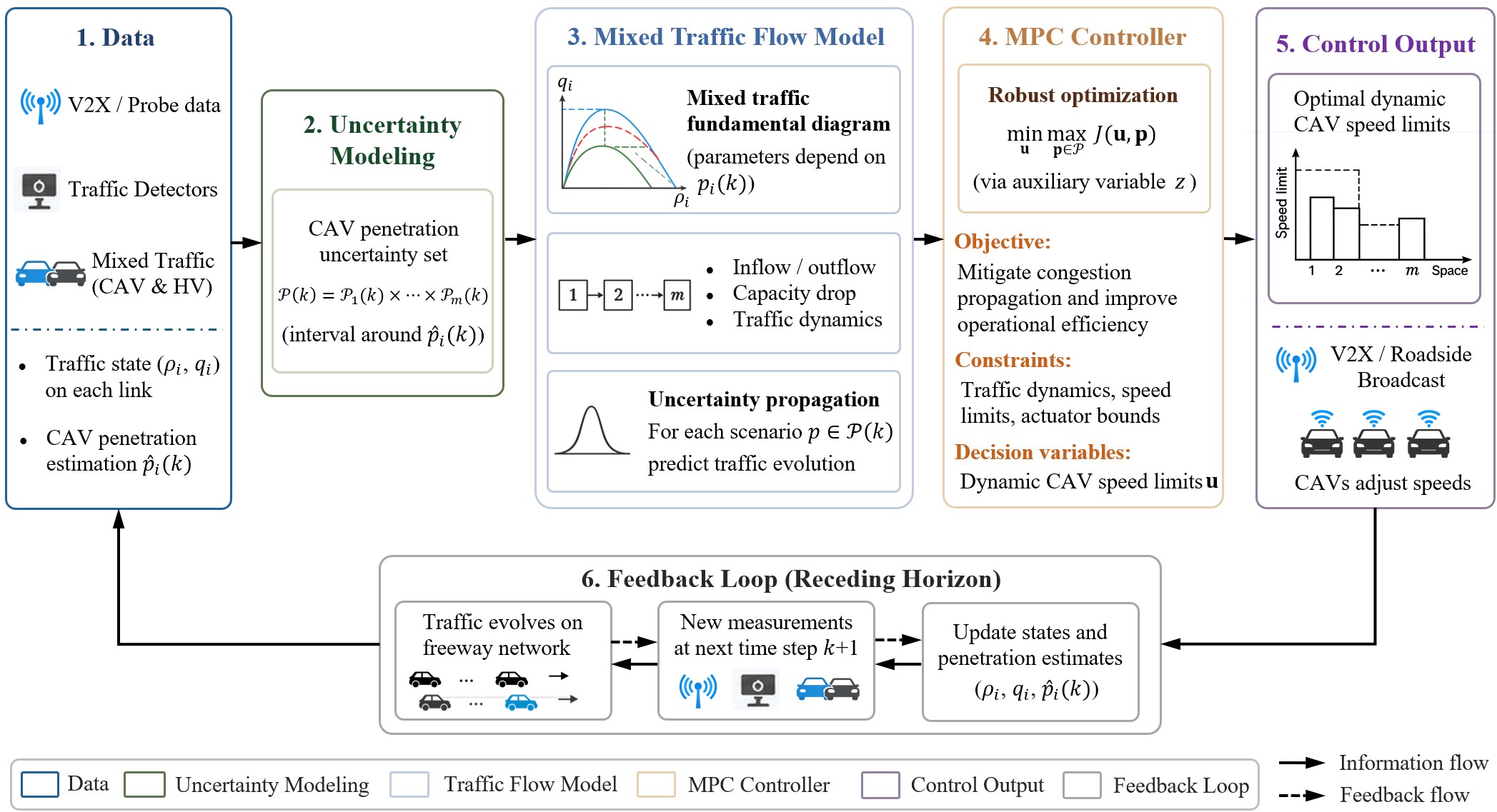}
\caption{Illustration of the proposed  model-based predictive control framework for dynamic speed limit control of CAVs in freeway networks considering traffic composition uncertainty in mixed traffic flow}
\label{framework}
\end{figure}

This study proposes a dynamic speed limit control framework for CAVs in freeway networks under uncertain mixed traffic conditions. The overall framework consists of three major modules, namely the mixed traffic flow modeling module, the uncertainty propagation module, and the robust model predictive controller, as illustrated in Figure \ref{framework}. At each control interval, the current traffic state, including density, flow, and estimated CAV penetration rate, are incorporated into an extended LTM, which predicts the evolution of traffic state over the prediction horizon. Since the CAV penetration rate cannot be perfectly observed in practice, the resulting traffic flow dynamics become uncertain. Such uncertainty propagates from the penetration estimation to the mixed fundamental diagram parameters and eventually affects the predicted states. Based on the predicted flow evolution, a robust model predictive controller determines the optimal CAV speed limits by minimizing network congestion while accounting for the worst-case uncertainty realization. The optimized speed limits are then applied to CAVs through V2I communication, forming a rolling optimization framework that continuously updates control actions as new measurements become available.

\subsection{Mixed traffic flow dynamics}\label{LTM_Formulation}
Consider a freeway network consisting of $N$ homogeneous links. Let $\rho_i(k)$, $q_i(k)$, $u_i(k)$, $p_i(k))$ denote the  density, flow, CAV speed limit, and CAV penetration rate of link $i$ at discrete time step $k$, respectively. The notations used in this study are summarized in Table \ref{notation}. The fundamental characteristic of the proposed model is that the macroscopic traffic parameters are no longer constant but dynamically depend on both the CAV penetration rate and the imposed CAV speed limits.

\begin{table*}[!t]
\captionsetup{font={small}}
\footnotesize
\renewcommand\arraystretch{1.08}
\caption{Descriptions of notations}
\label{notation}
\centering
\begin{tabularx}{\textwidth}{lX lX}
\toprule
\textbf{Notation} & \textbf{Definition} &
\textbf{Notation} & \textbf{Definition} \\
\midrule

$a_{\max}$ & Maximum allowable acceleration &
$p_i(k)$ & Actual CAV penetration rate of link $i$ \\

$\hat p_i(k)$ & Estimated CAV penetration rate &
$\Delta_i(k)$ & Penetration rate uncertainty bound \\

$\delta_i(k)$ & Penetration rate estimation error &
$\mathcal{P}_i(k)$ & Uncertainty set of link $i$ \\

$\mathcal{P}(k)$ & Network penetration uncertainty set &
$m$ & Number of freeway links \\

$L_i$ & Length of link $i$ &
$\Delta t$ & Sampling interval \\

$\rho_i(k)$ & Density of link $i$ &
$\rho_{j,i}$ & Jam density \\

$\rho_i^c(k)$ & Critical density &
$q_i(k)$ & Traffic flow \\

$q_i^{in}(k)$ & Link inflow &
$q_i^{out}(k)$ & Link outflow \\

$q_i^{dem}(k)$ & Sending demand &
$\bar q_i^{in}(k)$ & Maximum admissible inflow \\

$\bar q_i^{out}(k)$ & Mixed traffic capacity &
$C_i(k)$ & Mixed traffic capacity from headway \\

$C_i^{drop}(k)$ & Effective capacity considering capacity drop &
$\eta_i$ & Capacity drop ratio \\

$v_{f,i}^{mix}(k)$ & Mixed free-flow speed &
$v_f^H$ & Free-flow speed of HV \\

$v_f^C$ & Free-flow speed of CAV &
$w$ & Backward shockwave speed \\

$\bar h_i(k)$ & Average equilibrium headway &
$h^{HH}$ & HV--HV headway \\

$h^{HC}$ & HV--CAV headway &
$h^{CH}$ & CAV--HV headway \\

$h^{CC}$ & CAV--CAV headway &
$u_i(k)$ & CAV speed limit \\

$u_i^{\min}$ & Minimum speed limit &
$u_i^{\max}$ & Maximum speed limit \\

$\Delta u_i^{\max}$ & Maximum speed limit variation &
$\mathbf{u}(k)$ & Control vector \\

$\mathbf{s}(k)$ & Network state vector &
$s_i(k)$ & Number of vehicles stored on link $i$ \\

$N_i^{in}(k)$ & Cumulative inflow &
$N_i^{out}(k)$ & Cumulative outflow \\

$\bar N_i^{in}(k)$ & Maximum cumulative inflow &
$N_i^{out,f}(k)$ & Cumulative outflow under free flow\\

$N_i^{in,sh}(k)$ & Cumulative inflow constrained by shockwave &
$t_{f,i}(k)$ & Free-flow travel time \\

$t_{sh,i}$ & Shockwave travel time &
$n_{f,i}(k)$ & Discrete free-flow travel steps \\

$n_{sh,i}$ & Discrete shockwave travel steps &
$\gamma_{f,i}(k)$ & Free-flow interpolation coefficient \\

$\gamma_{sh,i}$ & Shockwave interpolation coefficient &
$d_i(k)$ & External traffic demand \\

$d_i^H(k)$ & HV demand &
$d_i^C(k)$ & CAV demand \\

$\Pi^{L}$ & Set of freeway links &
$\Pi^{O}$ & Set of origin links \\

$U_i$ & Upstream link set &
$T_p$ & Prediction horizon \\

$Q,R,S$ & Weighting matrices &
$\lambda_{\rho}$ & Capacity drop penalty weight \\

$\Psi(k)$ & Capacity drop risk &
$J(k)$ & Objective function \\

$D$ & Spatial difference operator &
$\Delta u(k)$ & Temporal speed limit variation \\

$N_s$ & Number of penetration scenarios &
$z$ & Auxiliary optimization variable \\

\bottomrule
\end{tabularx}
\end{table*}

\subsubsection{CAV penetration uncertainty}

Let $\hat{p}_i(k)$ denote the estimated CAV penetration rate of link $i$ at $k$. In practice, $\hat{p}_i(k)$ may differ from the actual penetration rate. In addition, communication delay and short-term fluctuations in vehicle composition may further introduce estimation errors. Therefore, the actual CAV penetration rate is modeled as
\begin{equation}
p_i(k)=\hat{p}_i(k)+\delta_i(k),
\end{equation}
where $\delta_i(k)$ is the estimation error. The error is assumed to be bounded:
\begin{equation}
-\Delta_i(k) \leq \delta_i(k) \leq \Delta_i(k).
\end{equation}

Equivalently,
\begin{equation}
p_i(k) \in \mathcal{P}_i(k)
=
\left[
\max\{0,\hat{p}_i(k)-\Delta_i(k)\},
\min\{1,\hat{p}_i(k)+\Delta_i(k)\}
\right].
\end{equation}

Different links may have different uncertainty levels because the number of observed CAVs and the traffic conditions are not the same across the freeway. For example, a link with fewer observed CAVs generally has a larger traffic composition uncertainty, while a link with more connected vehicle observations has a more reliable estimate. Thus, $\Delta_i(k)$ is allowed to vary across both space and time.

For the entire freeway network, the uncertainty set is defined  as the Cartesian product of the uncertainty sets associated with all links:
\begin{equation}
\mathcal{P}(k)
=
\mathcal{P}_1(k)
\times
\mathcal{P}_2(k)
\times
\cdots
\times
\mathcal{P}_m(k),
\label{uncertain set}
\end{equation}
where $m$ is the number of freeway links. This formulation keeps the uncertainty representation simple while allowing spatially heterogeneous penetration rate estimation errors. Note that at each control interval, the estimated penetration rate and its uncertainty bound are updated using the latest available observations. This enables the controller to adapt to time-varying CAV penetration and changing observation reliability.

\subsubsection{Mixed traffic fundamental diagram}
When designing speed limit control for CAVs in mixed traffic environments, estimating the traffic fundamental diagram (FD) is a crucial first step because the FD defines how speed control affects the macroscopic flow state. The FD depends on the speed and density. Let's first estimate the free-flow speed. Under free-flow conditions, interactions among vehicles are negligible and each vehicle travels at its desired speed. The vehicle free-flow speed can be regarded as an independent random variable. For a mixed traffic stream with a CAV penetration rate $p_i(k)$, the free-flow speed $V$ follows:
\begin{equation}
V=\left\{\begin{matrix}
 v_f^{H},~~~~ 1-p_i(k),\\
 \min(v_f^{C},u_i(k)),~~ p_i(k),
\end{matrix}\right.
\end{equation}
where $v_f^{H}$ and $v_f^{C}$ denote the desired free-flow speeds of HVs and CAVs, respectively. The operator guarantees that the actual CAV speed cannot exceed the imposed speed limit $u_i(k)$. Consequently, the expected free-flow speed is given by
\begin{equation}
\begin{split}
E(V)=(1-p_i(k))v_f^{H}+p_i(k)\min(v_f^{C},u_i(k)).
\end{split}
\end{equation}

The mixed free-flow speed equals the expectation of the individual desired speeds $v_{f,i}^{mix}=E(V)$. This formulation reflects that only the CAV component of the flow stream is directly influenced by the control action, while HVs continue to travel according to their own behavioral characteristics.  Based on the dynamically varying mixed free-flow speed, the mixed FD is formulated using a triangular relationship, as adopted in many existing studies \citep{zhang2025stochastic,makridis2023adaptive,zhou2020modeling}:
\begin{equation}
\begin{split}
q_i(k)=\min\left \{ v_{f,i}^{mix}\rho_i(k),w(\rho_{j,i}-\rho_i(k)) \right \} .
\end{split}
\end{equation}
where $w$ denotes the backward shockwave speed, and $\rho_{j,i}$ is the jam density. Capacity occurs where the free-flow branch and congested branch intersect, i.e., $v_{f,i}^{mix}\rho_i(k)=w(\rho_{j,i}-\rho_i(k))$, therefore the critical density $\rho_{i}^c(k)$ and capacity $\bar{q}_i^{out}(k)$ are:
\begin{equation}
    \rho_{i}^c(k)=\frac{w\rho_{j,i}}{v_{f,i}^{mix}+w}, 
\end{equation}
\begin{equation}
    \bar{q}_i^{out}(k)=\min\left \{C_i(k), \frac{v_{f,i}^{mix}w\rho_{j,i}}{v_{f,i}^{mix}+w} \right \}, 
\end{equation}
where $C_i(k)$ is the mixed capacity considering the impacts of heterogeneous headway. Four possible leader-follower interactions exist in mixed traffic: HV following HV (HH),
HV following CAV (HC), CAV following HV (CH), CAV following CAV (CC). Assuming random vehicle ordering, the average equilibrium headway can be expressed as:
\begin{equation}
\begin{split}
\bar{h}_i(k)
=\left(1-p_i(k)\right)^2 h_{HH}
+ p_i(k)\left(1-p_i(k)\right) h_{HC}
+ p_i(k)\left(1-p_i(k)\right) h_{CH}
+ p_i^2(k) h_{CC}.
\end{split}
\end{equation}

The corresponding mixed traffic capacity is
\begin{equation}
C_i(k)=\frac{3600}{\bar{h}_i(k)},
\end{equation}
where $h_{HH}$, $h_{HC}$, $h_{CH}$, and $h_{CC}$ denote the equilibrium time headways associated with different leader--follower combinations. Therefore, compared with traditional traffic models assuming a constant capacity, the proposed formulation enables the capacity to vary continuously with the CAV penetration rate and thus better reflects heterogeneous flow behavior.

\subsubsection{Link inflow dynamics}
For a freeway network, the cumulative inflow of each link is determined by either the discharge flow from its immediate upstream links or the external demand entering the network. Specifically, for a link \(i\), the cumulative inflow at time step \(k+1\) is updated as:
\begin{equation}
N_i^{{in}}(k+1)=
\begin{cases}
N_i^{{in}}(k)
+\displaystyle\sum_{j\in\mathcal U_i}
q_j^{{out}}(k)\Delta t,
&
i\in\Pi^{L}\cup\Pi^{O},
\\[2mm]
N_i^{{in}}(k)
+d_i(k)\Delta t,
&
i\in\Pi^{O},
\end{cases}
\label{eq:inflow_update}
\end{equation}
where \(\mathcal U_i\) denotes the set of links immediately upstream of link \(i\), \(d_i(k)\) is the external inflow demand, \(\Delta t\) represents the sampling interval, and $q_j^{{out}}(k)$ is the outflow rate of link $j$, which will be discussed later. $\Pi^{L}$ and $\Pi^{O}$ are the general links and origin links, respectively. In the mixed traffic environment, the external demand consists of both HVs and CAVs, i.e.,
\begin{equation}
d_i^{H}(k)=p_i(k)d_i(k),
\end{equation}
\begin{equation}
d_i^{C}(k)=(1-p_i(k))d_i(k).
\end{equation}

To avoid violating the physical storage limitation of each link, the cumulative inflow must satisfy:
\begin{equation}
N_i^{{in}}(k+1)
\le
\bar N_i^{{in}}(k+1),
\label{eq:in_limit}
\end{equation}
where \(\bar N_i^{{in}}(k+1)\) denotes the maximum admissible cumulative inflow. The upper bound of cumulative inflow is determined by the downstream traffic condition propagated through backward shockwaves \citep{wei2025link,hajiahmadi2015integrated}. The shockwave travel time is
\begin{equation}
t_{sh,i}
=
\frac{L_i}{w},
\end{equation}
where \(L_i\) is the length of link \(i\). The corresponding discrete propagation delay is
\begin{equation}
n_{sh,i}
=
\left\lfloor
\frac{t_{sh,i}}{\Delta t}
\right\rfloor ,
\end{equation}
and the interpolation coefficient is
\begin{equation}
\gamma_{sh,i}
=
n_{sh,i}
-
\frac{t_{sh,i}}{\Delta t},
\qquad
0\le\gamma_{sh,i}\le1.
\end{equation}

Accordingly, the cumulative inflow constrained by shockwave propagation can be estimated as
\begin{equation}
\begin{split}
N_i^{{in,sh}}(k+1)
=
\gamma_{sh,i}
N_i^{{out}}
(k+2-n_{sh,i})
+
(1-\gamma_{sh,i})
N_i^{{out}}
(k+1-n_{sh,i}),
\label{eq:shock}
\end{split}
\end{equation}
where \(N_i^{{out}}\) denotes the cumulative outflow of link \(i\).

The maximum admissible cumulative inflow is then obtained by incorporating the available storage space of the link,
\begin{equation}
\bar N_i^{{in}}(k+1)
=
N_i^{{in,sh}}(k+1)
+
\rho_{j,i}L_i.
\label{eq:maxin}
\end{equation}

The maximum allowable inflow rate of  \(i\) can be directly calculated from the cumulative inflow limit:
\begin{equation}
\bar q_i^{{in}}(k)
=
\frac{
\bar N_i^{{in}}(k)
-
\bar N_i^{{in}}(k-1)
}
{\Delta t}.
\label{eq:maxqin}
\end{equation}

The resulting inflow constraint provides a unified interface between the mixed traffic flow model and the subsequent dynamic speed limit control strategy, ensuring that the predicted traffic evolution always satisfies the physical storage limitation of each freeway link.
\subsubsection{Link outflow dynamics}
After determining the admissible inflow of each link, the link outflow is further formulated by considering both the mixed traffic discharge capacity and the capacity drop effect. For link \(i\), the cumulative outflow at time step \(k+1\) is updated as
\begin{equation}
N_i^{{out}}(k+1)
=
N_i^{{out}}(k)
+
q_i^{{out}}(k)\Delta t ,
\label{eq:outflow_update}
\end{equation}
where \(q_i^{{out}}(k)\) is the outflow rate of link \(i\) at time step \(k\).

The outflow of a link is constrained by three factors: the number of vehicles that have completed their travel within the link, the mixed capacity of the link, and the maximum admissible inflow of the downstream link. Therefore, the outflow rate is given by
\begin{equation}
q_i^{{out}}(k)
=
\min
\left\{
q_i^{{dem}}(k),
C_i^{{drop}}(k),
\bar q_{i+1}^{{in}}(k)
\right\},
\label{eq:outflow_rate}
\end{equation}
where \(q_i^{{dem}}(k)\) is the sending demand of link \(i\), \(C_i^{{drop}}(k)\) is the effective mixed capacity after considering the capacity drop effect, and \(\bar q_{i+1}^{{in}}(k)\) is the maximum allowable inflow rate of the downstream link.

The sending demand is determined by the free-flow propagation of vehicles within the link. Let
\begin{equation}
t_{f,i}(k)
=
\frac{L_i}{v_{f,i}^{{mix}}(k)}
\end{equation}
denote the free-flow travel time of link \(i\). The corresponding discrete travel delay is
\begin{equation}
n_{f,i}(k)
=
\left\lfloor
\frac{t_{f,i}(k)}{\Delta t}
\right\rfloor ,
\end{equation}
and the interpolation coefficient is
\begin{equation}
\gamma_{f,i}(k)
=
n_{f,i}(k)
-
\frac{t_{f,i}(k)}{\Delta t}.
\end{equation}

Thus, the maximum cumulative outflow determined by free-flow propagation can be estimated as
\begin{equation}
\begin{split}
N_i^{{out,f}}(k+1)
=
\gamma_{f,i}(k)
N_i^{{in}}
(k+2-n_{f,i}(k))
+\left(1-\gamma_{f,i}(k)\right)
N_i^{{in}}
(k+1-n_{f,i}(k)).
\label{eq:ff_out}
\end{split}
\end{equation}

The corresponding sending demand is
\begin{equation}
q_i^{{dem}}(k)
=
\frac{
N_i^{{out,f}}(k+1)
-
N_i^{{out}}(k)
}
{\Delta t}.
\label{eq:sending_demand}
\end{equation}

For mixed traffic, the nominal capacity is determined by the average equilibrium headway among different leader-follower vehicle combinations. Under the random mixing assumption, the nominal mixed capacity is expressed by Eq. (\ref{eq:capacity_drop}). To capture the reduced discharge rate after congestion formation, the capacity drop effect is incorporated into the mixed capacity. If the critical density is \(\rho_i^{c}(k)\), the effective capacity is formulated as
\begin{equation}
C_i^{{drop}}(k)
=
\begin{cases}
C_i^{{mix}}(k),
&
\rho_i(k)\le \rho_i^{c}(k),
\\[1mm]
\left(1-\eta_i\right)C_i^{{mix}}(k),
&
\rho_i(k)> \rho_i^{c}(k),
\end{cases}
\label{eq:capacity_drop}
\end{equation}
where \(\eta_i\in[0,1)\) is the capacity drop ratio. A larger \(\eta_i\) indicates a stronger reduction in the bottleneck discharge capacity after the link becomes congested.

The density of link \(i\) is calculated from the difference between cumulative inflow and outflow as
\begin{equation}
\rho_i(k)
=
\frac{
N_i^{{in}}(k)
-
N_i^{{out}}(k)
}
{L_i}.
\label{eq:density_cumulative}
\end{equation}

Combining Eqs.~\eqref{eq:outflow_rate}--\eqref{eq:capacity_drop}, the outflow dynamics can be compactly written as
\begin{equation}
\begin{split}
q_i^{{out}}(k)
=\min
\left\{
q_i^{{dem}}(k),
C_i^{{mix}}(k)\left[1-\eta_i \mathbb{I}\left(\rho_i(k)>\rho_i^{c}(k)\right)\right],
\bar q_{i+1}^{{in}}(k)
\right\},
\label{eq:compact_outflow}
\end{split}
\end{equation}
where \(\mathbb{I}(\cdot)\) is an indicator function. Eq.~\eqref{eq:compact_outflow} shows that the link outflow is jointly governed by vehicle propagation, mixed traffic capacity, capacity drop, and downstream supply. In this formulation, the CAV speed limit affects the outflow dynamics through the mixed free-flow speed and mixed capacity, while the capacity drop mechanism captures the reduction in discharge flow once the bottleneck becomes congested.
\subsection{Controller formulation}
\label{Controller formulation}
\begin{figure}
\captionsetup{font={small}}
\centering
\includegraphics[width=6.2in]{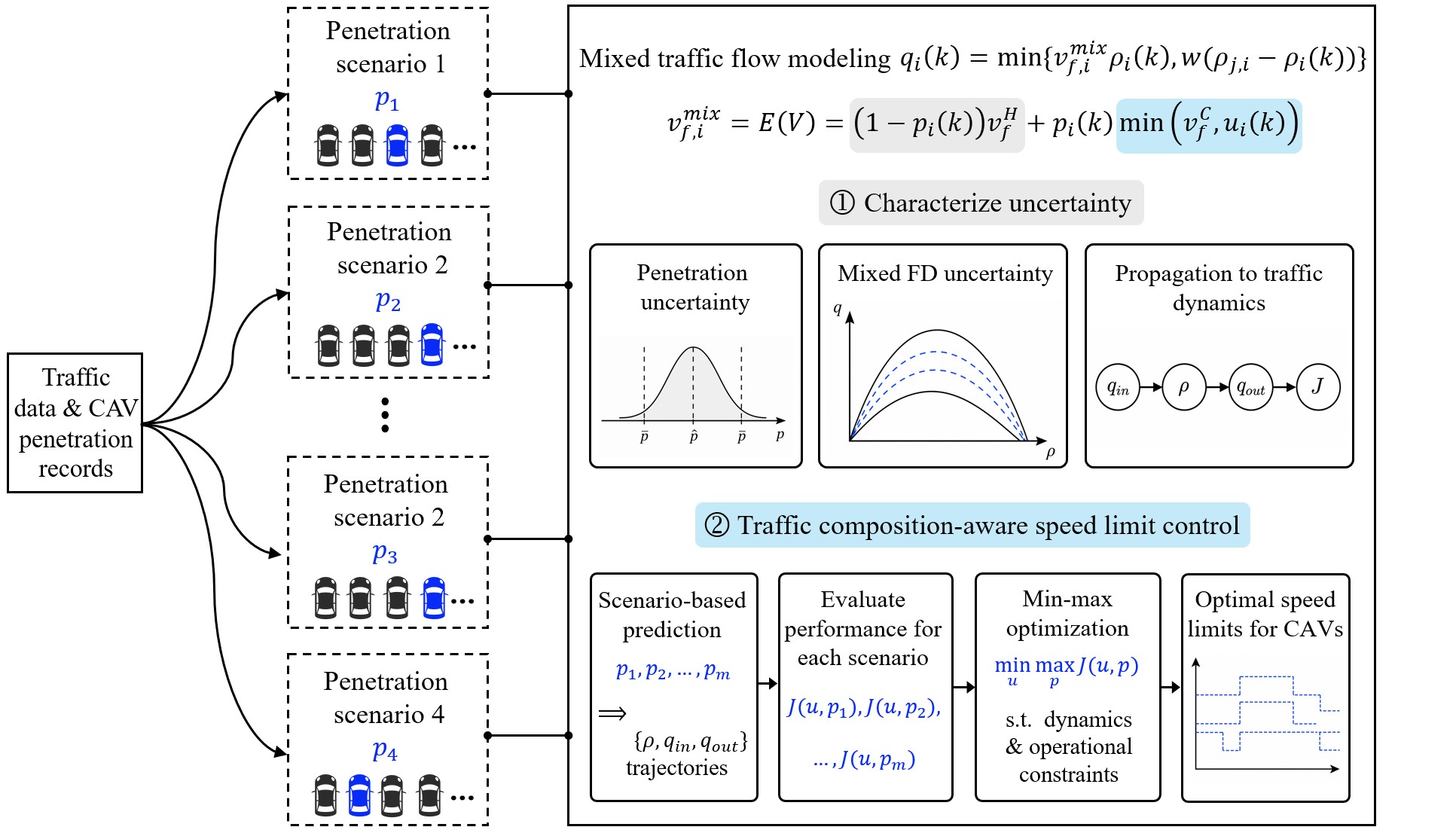}
\caption{Overview of the proposed dynamic CAV speed limit control framework with penetration uncertainty propagation in mixed traffic flow, where a traffic composition-aware MPC optimizes CAV speed limits.}
\label{framework2}
\end{figure}

Based on the mixed traffic flow dynamics formulated above, this section develops a dynamic speed limit controller for CAVs in freeway networks. The overview of the proposed framework is shown in Figure \ref{framework2}. The objective of the controller is to determine the optimal CAV speed limits over a finite prediction horizon such that freeway congestion and capacity drop risks are mitigated while avoiding excessive temporal and spatial variations in speed control actions.

Let \(T_p\) denote the prediction horizon. At each control time step \(k\), the controller optimizes the CAV speed limits over the horizon as
\begin{equation}
\mathbf{u}(k)
=
\left[
u(k|k),u(k+1|k),\ldots,u(k+T_p-1|k)
\right],
\end{equation}
where
\begin{equation}
u(k+t|k)
=
\left[
u_1(k+t|k),
u_2(k+t|k),
\ldots,
u_m(k+t|k)
\right]^{\top}
\end{equation}
is the vector of CAV speed limits for all freeway links at prediction step \(k+t\), and \(m\) is the number of links in the network.

The system state is defined by the number of vehicles stored on each link, which can be calculated from cumulative inflow and outflow as
\begin{equation}
s_i(k)
=
N_i^{{in}}(k)
-
N_i^{{out}}(k).
\end{equation}

Thus, the network state vector is
\begin{equation}
\mathbf{s}(k)
=
\left[
s_1(k),s_2(k),\ldots,s_m(k)
\right]^{\top}.
\end{equation}

The state transition is governed by the mixed traffic link dynamics:
\begin{equation}
s_i(k+1)
=
s_i(k)
+
\left(
q_i^{{in}}(k)
-
q_i^{{out}}(k)
\right)\Delta t,
\label{eq:state_transition}
\end{equation}
where \(q_i^{{in}}(k)\) and \(q_i^{{out}}(k)\) are determined by the cumulative inflow and outflow dynamics defined in the previous subsections.

The objective function consists of three parts. The first term minimizes the total number of vehicles accumulated in the network, which is equivalent to reducing the total travel time spent. The second term penalizes abrupt temporal changes of CAV speed limits between consecutive control intervals. The third term penalizes large spatial differences in speed limits between adjacent links to maintain smooth speed transitions along the freeway. The cost function is formulated as:
\begin{equation}
\begin{split}
J(k)
=
\sum_{t=1}^{T_p}
\left\|
\mathbf{s}(k+t|k)
\right\|_{Q}^{2}
+
\sum_{t=0}^{T_p-1}
\left\|
\Delta \mathbf{u}(k+t|k)
\right\|_{R}^{2}
+\sum_{t=0}^{T_p-1}
\left\|
D\mathbf{u}(k+t|k)
\right\|_{S}^{2},
\label{eq:mpc_cost}
\end{split}
\end{equation}
where \(Q\), \(R\), and \(S\) are positive semi-definite weighting matrices. The temporal variation of the speed control action is defined as:
\begin{equation}
\Delta \mathbf{u}(k+t|k)
=
\mathbf{u}(k+t|k)
-
\mathbf{u}(k+t-1|k),
\end{equation}
where \(\mathbf{u}(k-1|k)\) is the speed limit implemented at the previous control interval when \(t=0\). The matrix \(D\) is a spatial difference operator, and \(D\mathbf{u}(k+t|k)\) represents the speed differences between neighboring freeway links.

To explicitly account for capacity drop effect, an additional risk penalty is introduced when predicted density exceeds the critical value. The capacity drop risk is defined as:
\begin{equation}
\Psi(k+t|k)
=
\sum_{i=1}^{m}
\left[
\max
\left(
0,
\rho_i(k+t|k)-\rho_i^{c}
\right)
\right]^2.
\end{equation}

The objective function can then be extended as
\begin{equation}
\begin{split}
J(k)
=
\sum_{t=1}^{T_p}
\left\|
\mathbf{s}(k+t|k)
\right\|_{Q}^{2}
+
\sum_{t=0}^{T_p-1}
\left\|
\Delta \mathbf{u}(k+t|k)
\right\|_{R}^{2}
+\sum_{t=0}^{T_p-1}
\left\|
D\mathbf{u}(k+t|k)
\right\|_{S}^{2}
+
\lambda_{\rho}
\sum_{t=1}^{T_p}
\Psi(k+t|k),
\label{eq:mpc_cost_cd}
\end{split}
\end{equation}
where \(\lambda_{\rho}\) is the weighting coefficient associated with the capacity drop risk.

Since the CAV penetration rate is not perfectly known in practice, the uncertainty set is defined as:
\begin{equation}
\mathcal{P}_i(k)
=
\left\{
p_i(k):
\hat p_i(k)-\Delta_i^p
\le
p_i(k)
\le
\hat p_i(k)+\Delta_i^p
\right\}.
\end{equation}

For the entire freeway network, the admissible uncertainty set is constructed as shown in Eq. (\ref{uncertain set}). The robust dynamic speed control problem is thus formulated as a min--max optimization problem:
\begin{equation}
\min_{\mathbf{u}(k)}
\;
\max_{\mathbf{p}(k+t|k)\in\mathcal{P}(k+t)}
J(k),
\label{eq:robust_mpc}
\end{equation}
subject to the mixed traffic flow dynamics in Eqs.~\eqref{eq:inflow_update}, \eqref{eq:in_limit}, \eqref{eq:outflow_update}, and the following operational constraints:
\begin{equation}
u_i^{\min}
\le
u_i(k+t|k)
\le
u_i^{\max},
\quad
\forall i,\; t,
\label{eq:u_bound}
\end{equation}
\begin{equation}
\left|
u_i(k+t|k)
-
u_i(k+t-1|k)
\right|
\le
\Delta u_i^{\max},
\quad
\forall i,\; t,
\label{eq:u_rate}
\end{equation}
\begin{equation}
\left|
u_i(k+t|k)
-
u_{i+1}(k+t|k)
\right|
\le
\Delta u_{s}^{\max},
\quad
\forall i,\; t,
\label{eq:u_space}
\end{equation}
\begin{equation}
0
\le
\rho_i(k+t|k)
\le
\rho_{j,i},
\quad
\forall i,\; t.
\label{eq:rho_bound}
\end{equation}
Eq.~\eqref{eq:u_bound} ensures that the optimized speed limits remain within the allowable speed range. Eq.~\eqref{eq:u_rate} prevents abrupt temporal variations in CAV speed limits, while Eq.~\eqref{eq:u_space} guarantees smooth spatial transitions between adjacent links. Eq.~\eqref{eq:rho_bound} enforces the physical density bound of each link.

At each control interval, only the first optimized control action is implemented:
\begin{equation}
\mathbf{u}^{*}(k)
=
\mathbf{u}^{*}(k|k).
\end{equation}

Then, the traffic states are updated using newly observed measurements, and the optimization problem is solved again at the next time step. This receding-horizon implementation enables the controller to continuously adapt to the evolving mixed traffic conditions and traffic composition uncertainty.

Note that the proposed formulation differs from deterministic speed control methods, which optimize CAV speed limits based on an assumed accurate estimate of the penetration rate. Instead, it searches for a control strategy that performs satisfactorily across all admissible realizations of mixed traffic flow. As a result, the derived speed control strategy is more robust to uncertainties and estimation errors in CAV penetration rates.

\subsection{Solution method}
\label{Solution method}
The optimization problem formulated in Eq.~\eqref{eq:robust_mpc} is a nonlinear min--max optimization problem, in which the CAV speed limits influence the mixed traffic parameters, flow propagation, and capacity drop dynamics simultaneously. Moreover, the uncertain CAV penetration rates introduce additional nonconvexity into the prediction model, making it difficult to derive a closed-form solution.

To improve computational tractability, the proposed robust optimization problem is solved using a scenario-based framework as summarized by Algorithm \ref{alg:robust_mpc}. Instead of optimizing against infinitely many uncertainty realizations, the admissible penetration uncertainty set is discretized into a finite number of representative scenarios,
\begin{equation}
\mathcal{P}
=
\left\{
\mathbf{p}^{(1)},
\mathbf{p}^{(2)},
\ldots,
\mathbf{p}^{(N_s)}
\right\},
\end{equation}
where \(N_s\) denotes the number of penetration scenarios.

For each scenario, the mixed traffic parameters, including the free-flow speed and mixed capacity, are updated according to the corresponding penetration rate, and the traffic evolution is predicted using the proposed extended LTM. The objective value associated with scenario \(s\) is denoted by
\begin{equation}
J^{(s)}(\mathbf{u}),
\qquad
s=1,\ldots,N_s.
\end{equation}

The original min--max problem
\begin{equation}
\min_{\mathbf{u}}
\;
\max_{\mathbf{p}\in\mathcal{P}}
J(\mathbf{u},\mathbf{p})
\end{equation}
can therefore be approximated by
\begin{equation}
\min_{\mathbf{u}}
\;
\max_{s=1,\ldots,N_s}
J^{(s)}(\mathbf{u}),
\label{eq:scenario_problem}
\end{equation}
which transforms the continuous uncertainty into a finite-dimensional optimization problem.

To further improve numerical efficiency, an auxiliary variable \(z\) is introduced to replace the maximum operator. The optimization problem is equivalently reformulated as
\begin{equation}
\min_{\mathbf{u},z}
\quad
z
\end{equation}
subject to
\begin{equation}
J^{(s)}(\mathbf{u})
\le
z,
\qquad
s=1,\ldots,N_s,
\label{eq:epigraph}
\end{equation}
together with the dynamic constraints and speed limit constraints presented in the previous subsection. It can be efficiently solved using sequential quadratic programming. 

\begin{algorithm}[!t]
\caption{Traffic composition-aware MPC for CAV speed limit control}
\label{alg:robust_mpc}
\begin{algorithmic}[1]
\Require Initial state $x(0)$, horizon $T_p$, control interval $\Delta t_c$, speed bounds $u^{\min}$ and $u^{\max}$, parameters $\theta$, weighting matrices $Q,R,S$, and capacity drop weight $\lambda_\rho$
\Ensure Implemented CAV speed limits $u(k)$

\For{each control time step $k$}
    \State Observe current traffic state $x(k)$
    \State Estimate CAV penetration rate $\hat p(k)$ and uncertainty bound $\Delta(k)$
    \State Construct uncertainty set $\mathcal{P}(k)=[\hat p(k)-\Delta(k),\hat p(k)+\Delta(k)]$
    \State Generate scenarios $\mathcal{S}(k)=\{p^{(1)}(k),p^{(2)}(k),\ldots,p^{(N_s)}(k)\}$
    \State Initialize $J_{\mathrm{best}}\leftarrow +\infty$

    \For{each candidate speed limit sequence $\mathbf{u}(k)$}
        \State Initialize worst case $J_{\mathrm{wc}}\leftarrow -\infty$

        \For{each scenario $p^{(s)}(k)\in\mathcal{S}(k)$}
            \State Set $x^{(s)}(k|k)\leftarrow x(k)$

            \For{$t=0$ to $T_p-1$}
                \State Update $v_{f,i}^{\mathrm{mix}}(k+t|k)$, $C_i^{\mathrm{mix}}(k+t|k)$, and $\rho_{c,i}(k+t|k)$
                \State Predict state:
                \State $x^{(s)}(k+t+1|k)=f\bigl(x^{(s)}(k+t|k),u(k+t|k),p^{(s)}(k)\bigr)$
            \EndFor

            \State Compute objective:
            \State $J^{(s)}(\mathbf{u}(k))=J\bigl(x^{(s)}(k+1:k+T_p|k),\mathbf{u}(k)\bigr)$
            \State Update $J_{\mathrm{wc}}\leftarrow \max\{J_{\mathrm{wc}},J^{(s)}(\mathbf{u}(k))\}$
        \EndFor

        \If{$J_{\mathrm{wc}}<J_{\mathrm{best}}$}
            \State $J_{\mathrm{best}}\leftarrow J_{\mathrm{wc}}$
            \State $\mathbf{u}^*(k)\leftarrow \mathbf{u}(k)$
        \EndIf
    \EndFor

    \State Implement the first optimized speed limit $u(k)\leftarrow u^*(k|k)$
    \State Move to the next control interval and update measurements
\EndFor

\end{algorithmic}
\end{algorithm}



\section{Case study}
\label{simulation} 
This section presents numerical simulation results to evaluate the effectiveness of the proposed MPC framework under different traffic conditions. Two case studies are conducted. The first considers a single-bottleneck freeway corridor. The second examines a larger multi-bottleneck freeway network with merge-diverge spillback, which is used to evaluate the robustness and scalability of the proposed method and to benchmark its performance against existing control strategies.
\subsection{Single-bottleneck freeway corridor}
\begin{figure}[H]
\captionsetup{font={small}}
\centering
\includegraphics[width=6.3in]{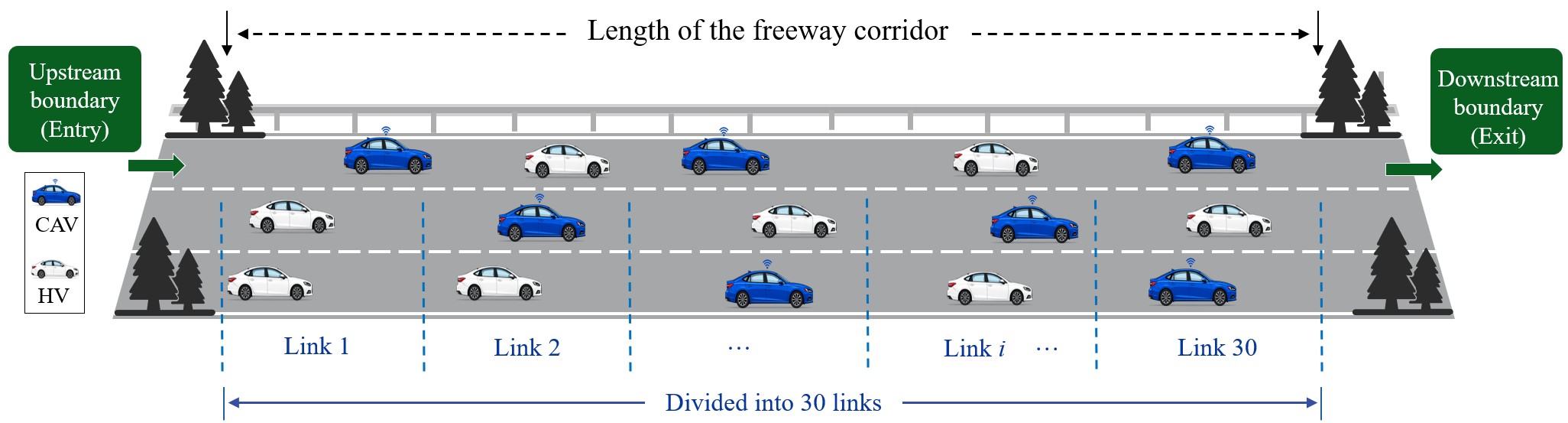}
\vspace{-5pt}
\caption{Schematic illustration of the freeway corridor used in the case study. The corridor is divided into 30 links for traffic flow modeling and control. Mixed traffic consisting of CAVs and HVs enters from the upstream boundary. }
\label{freeway}
\end{figure}

\begin{figure}[H]
\captionsetup{font={small}}
\centering
\includegraphics[width=3.1in]{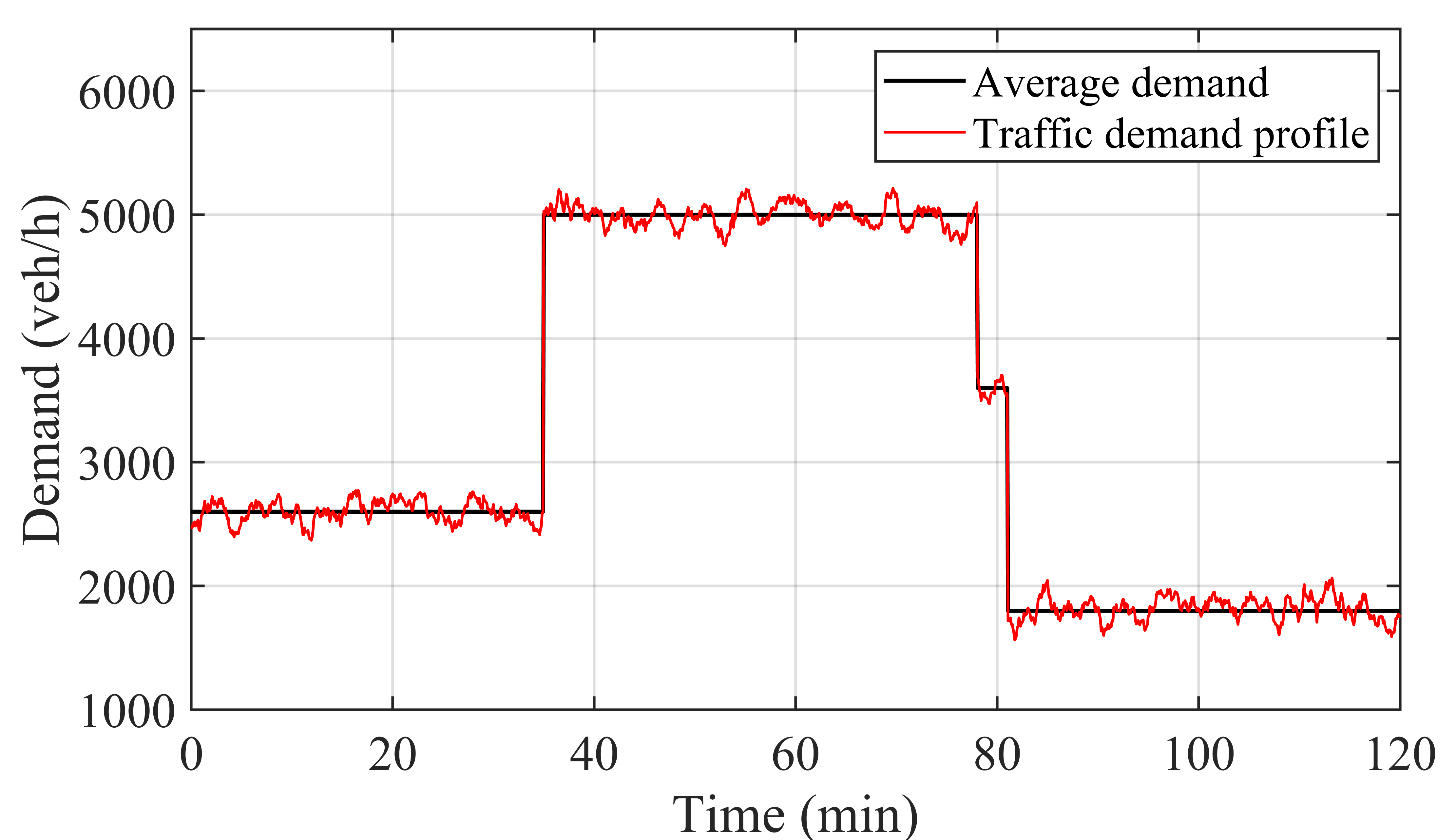}
\vspace{-5pt}
\caption{Illustration of the traffic demands in the simulated freeway corridor}
\label{demand_case1}
\end{figure}

The first case study is designed to examine the basic control mechanism of the proposed MPC in a single recurrent bottleneck corridor. This corridor provides a clearer setting for interpreting how traffic composition affects the resulting speed control decisions. The corridor has a total length of 7.5 km and is discretized into 30 links with a spatial resolution of 0.25 km as shown in Figure \ref{freeway}. The traffic evolution is simulated with a sampling interval of 5 s. A recurrent bottleneck is located at the downstream end of the corridor, where capacity drop occurs once the local density exceeds a predefined threshold. The traffic demand profile is designed to reproduce a typical peak-hour congestion scenario as shown in Figure \ref{demand_case1}. During the peak period, the upstream demand exceeds the bottleneck discharge capacity, resulting in congestion formation and backward shockwave propagation. 

Three control strategies are considered for comparison: 1) No control, where no speed regulation is applied and vehicles operate under prevailing traffic conditions; 2) Baseline MPC (MPC-B), which represents a conventional MPC approach widely adopted in the literature, assumes a deterministic nominal CAV penetration rate throughout the prediction horizon \citep{han2021linear}. 3) Proposed MPC (MPC-P), which explicitly incorporates uncertainty in CAV penetration rates into the flow prediction and optimization process. 

The detailed CAV penetration rate settings are presented in Table \ref{tab:case1_penetration}. The nominal CAV penetration rate is set to 35\%, representing an intermediate mixed-traffic condition in which both CAVs and HVs coexist. The actual penetration rate is assumed to lie within a bounded uncertainty interval around the nominal estimate $p_i(k)\in [\hat p_i(k)-\Delta_i,\hat p_i(k)+\Delta_i]$. Specifically, the uncertainty bound is $\Delta_i=0.1$. Thus, the admissible penetration rate interval is $p_i(k)\in[0.25,0.45]$. The MPC-B uses only the nominal penetration rate $\hat p_i(k)=0.35$ in prediction, whereas the proposed MPC-P evaluates traffic evolution under representative penetration rate scenarios within $[0.25,0.45]$.

\begin{figure}[H]
\captionsetup{font={small}}
\centering  
\subfigure[Speed with no control]{
\includegraphics[width=2.4in]{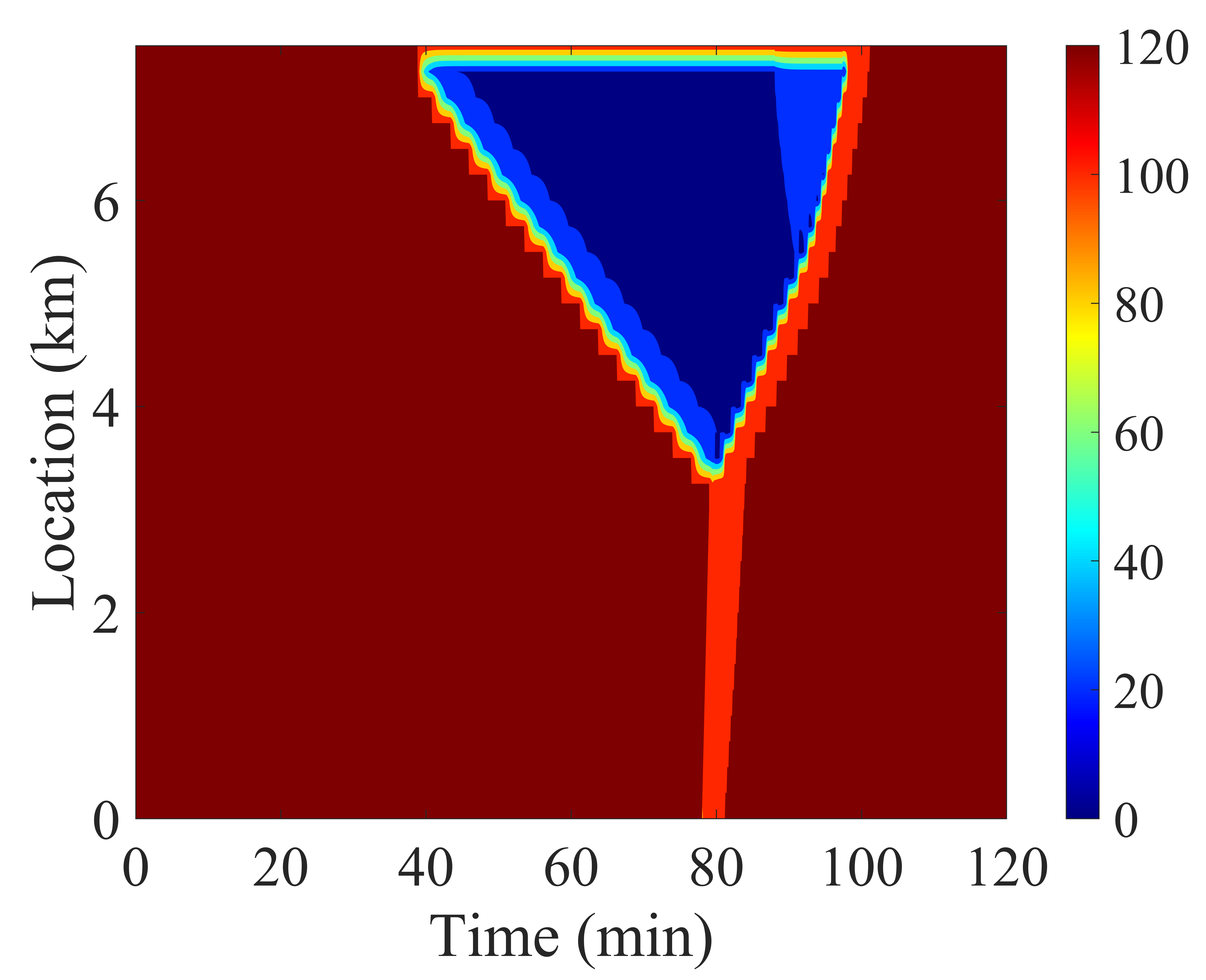}}~~~\subfigure[Flow with no control]{
\includegraphics[width=2.4in]{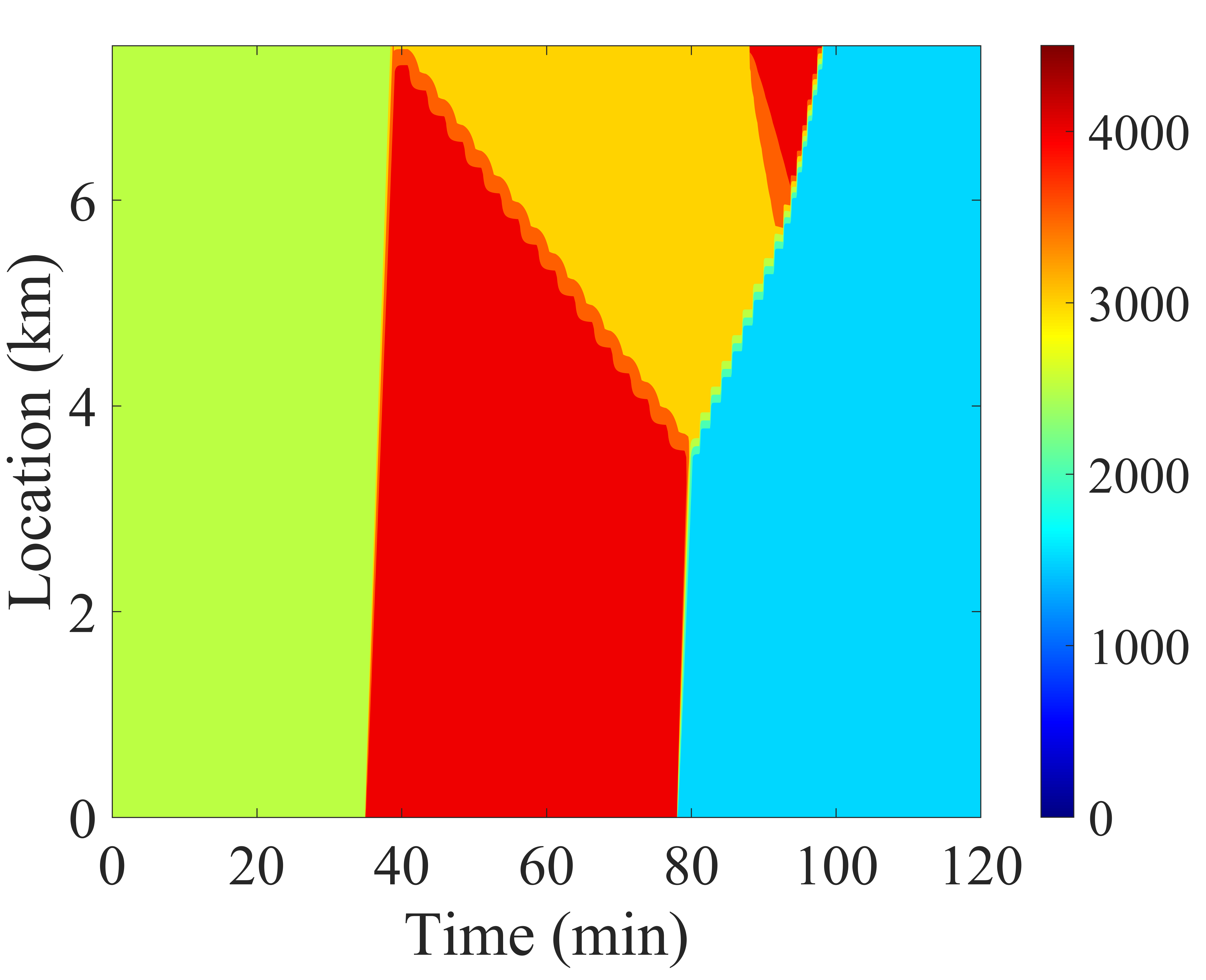}}
\subfigure[Speed with MPC-B]{
\includegraphics[width=2.4in]{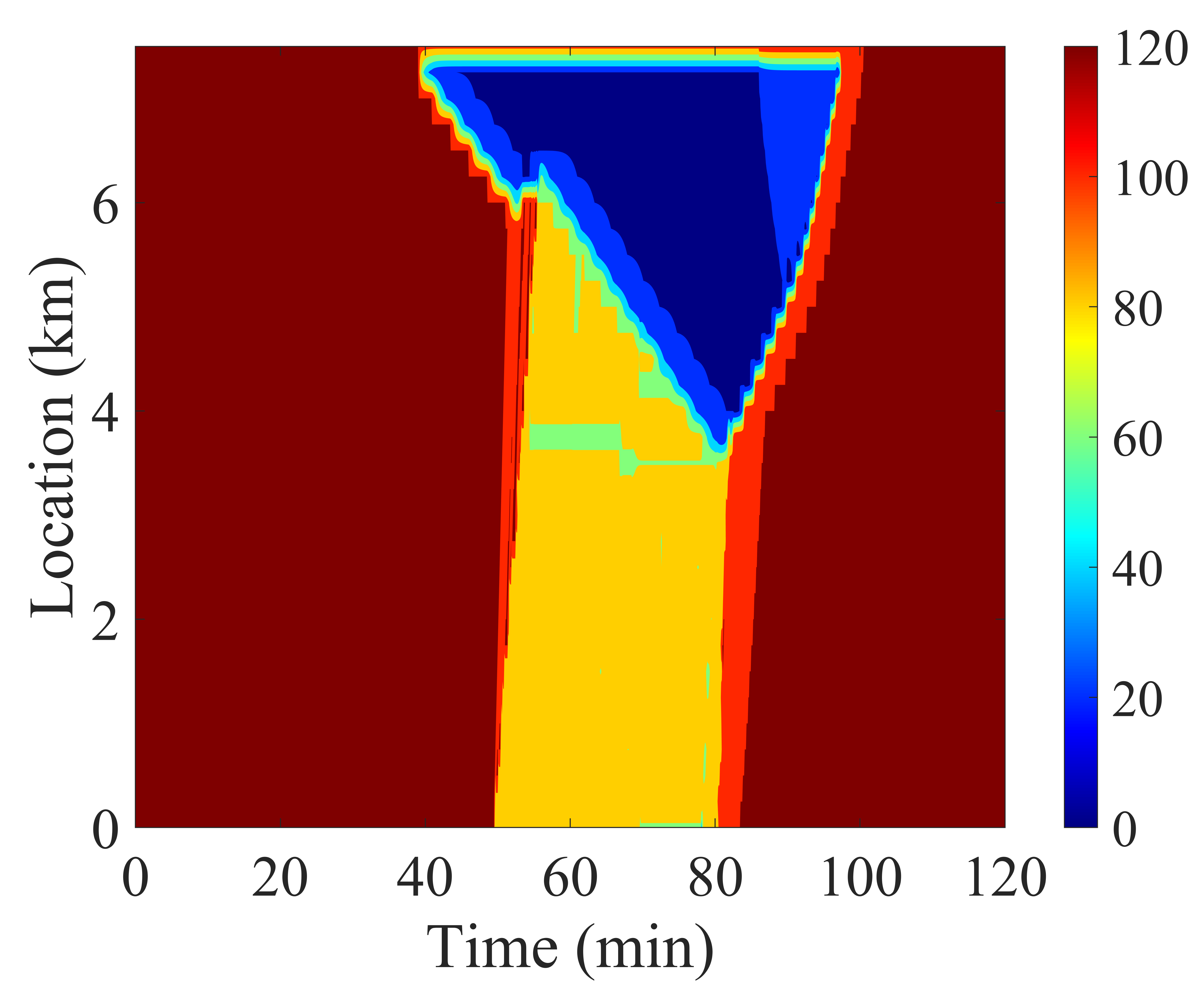}}~~~\subfigure[Flow with MPC-B]{
\includegraphics[width=2.4in]{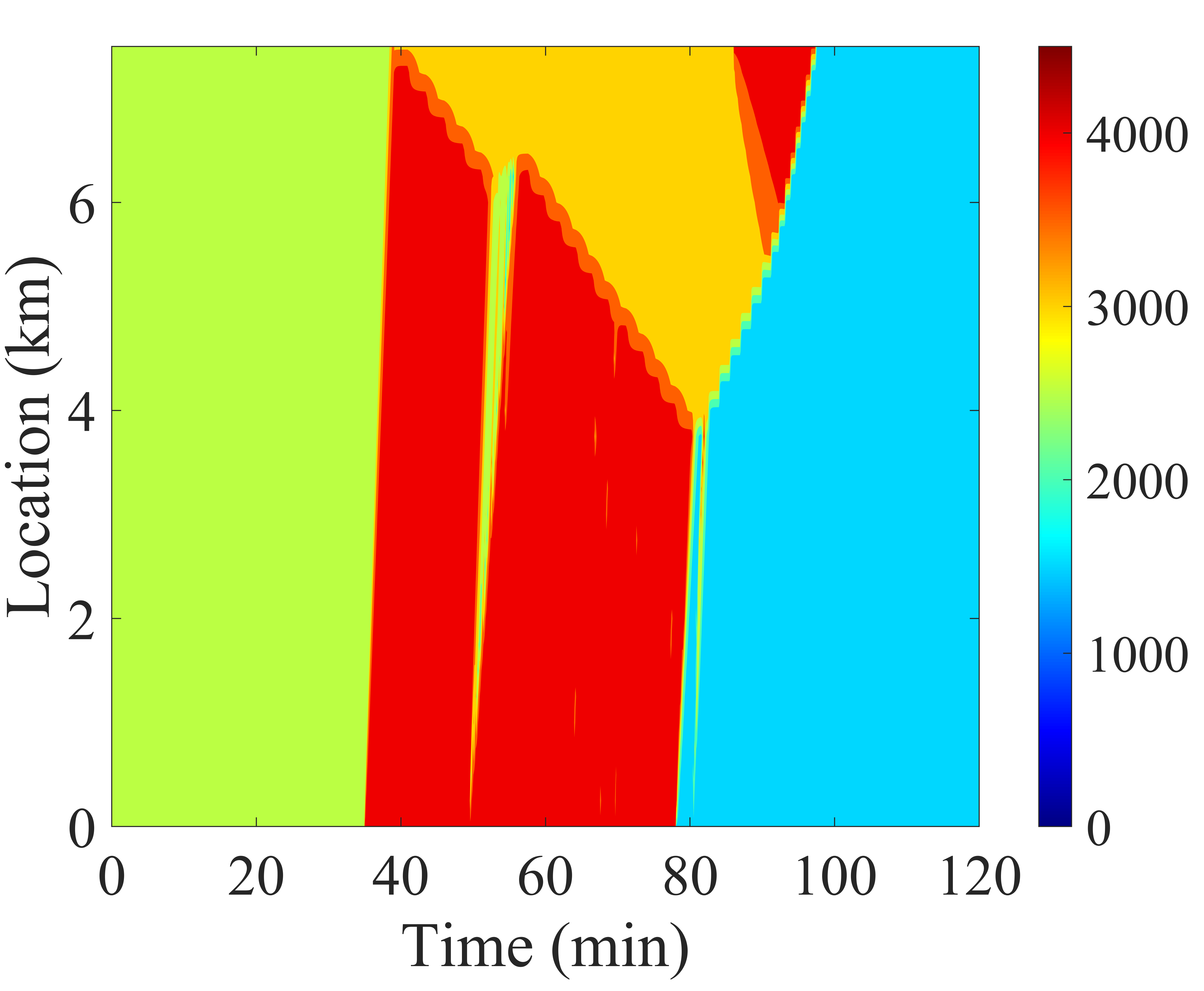}}
\subfigure[Speed with MPC-P]{
\includegraphics[width=2.4in]{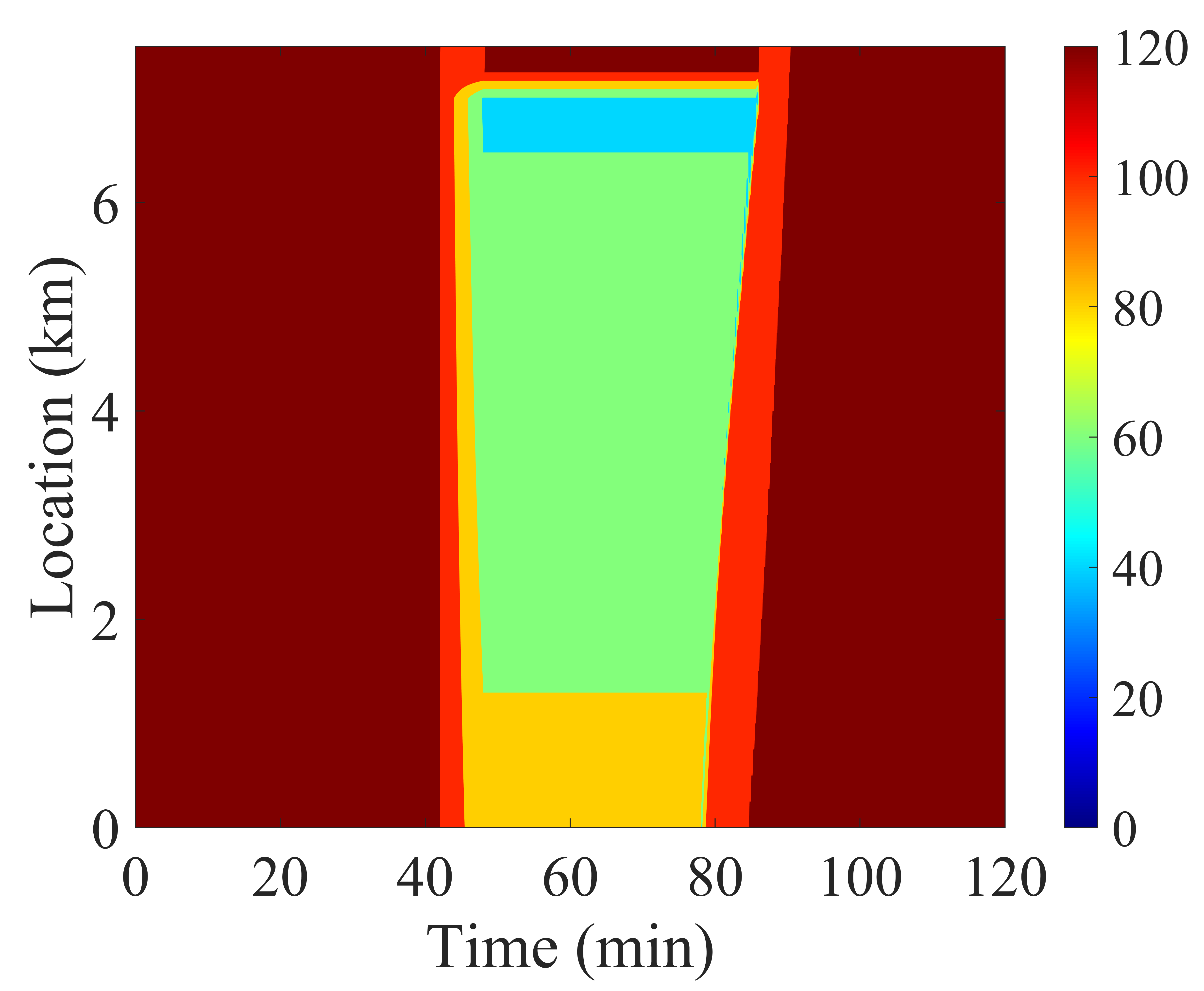}}~~~\subfigure[Flow with MPC-P]{
\includegraphics[width=2.4in]{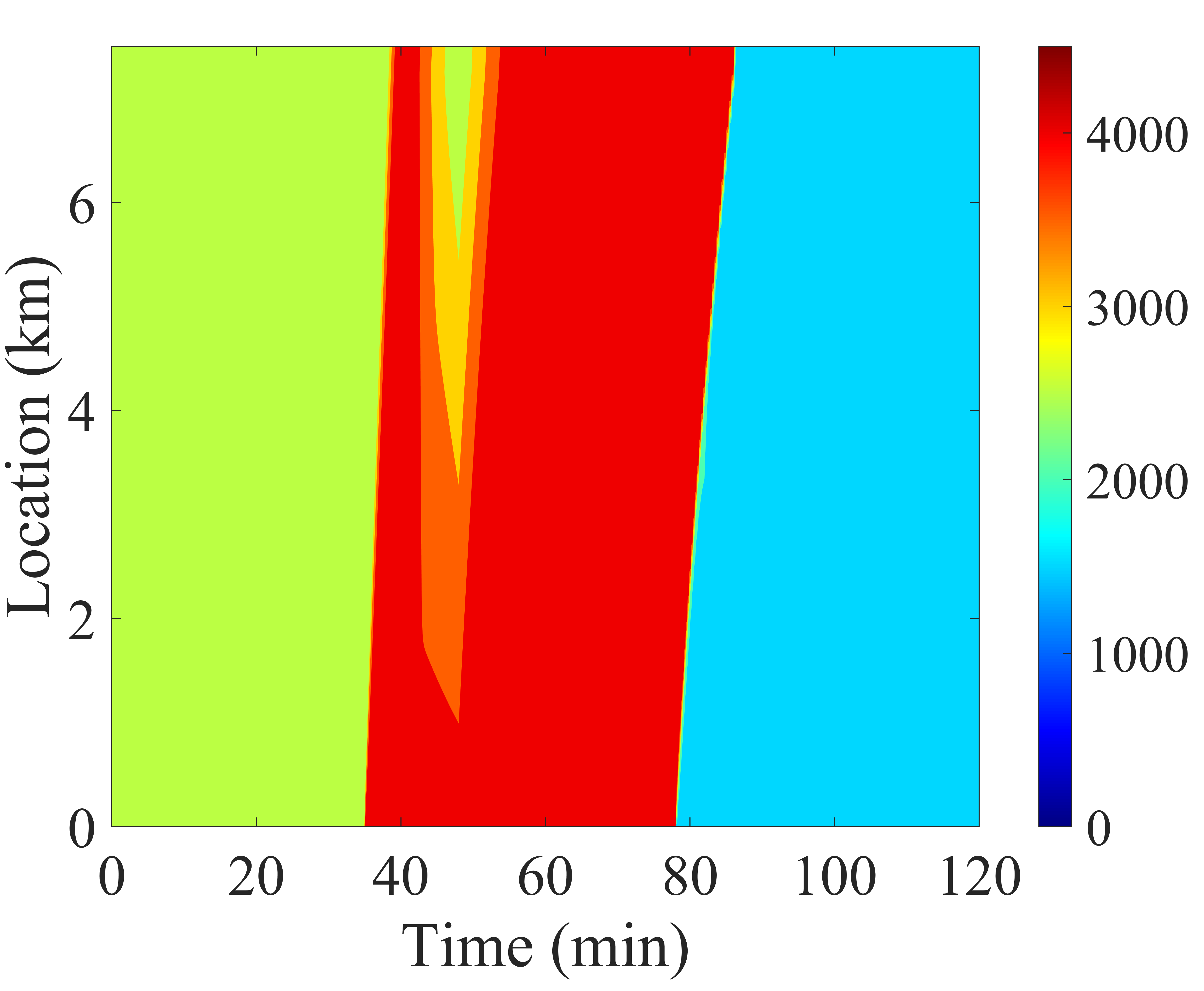}}
\vspace{-5pt}
\caption{Comparison of spatiotemporal average speed (km/h) and flow (veh/h) patterns under different freeway control strategies}
\label{c1density}
\end{figure}

The traffic dynamics are simulated using a macroscopic cell transmission model integrated with a supply-proportional node model to assess the effectiveness of the proposed MPC-P and the baseline MPC-B. This process model is different from the extended LTM adopted within the MPC-P and MPC-B, but it maintains a high level of computational efficiency and providing an independent fair evaluation. The calibrated model parameters are shown in Table \ref{freeway parameters}, where the average headway is adopted from the calibrated results reported by \cite{hussain2016freeway}. To balance prediction accuracy and computational burden, the prediction horizon  $T_p$ is set to 500 s in this case. The spatiotemporal speed and flow distributions under the three control strategies are illustrated in Figure~\ref{c1density}.

\vspace{-5pt}

\vspace{-5pt}
\begin{table}[H]
\captionsetup{font={small}}
\footnotesize
\centering
\caption{CAV penetration rate settings}
\vspace{-5pt}
\label{tab:case1_penetration}
\begin{tabular}{l c}
\hline
Item & Value \\
\hline
Nominal penetration rate $\hat p_i$ & 0.35 \\
Uncertainty bound $\Delta_i$ & 0.10 \\
Lower penetration scenario & 0.25 \\
Nominal penetration scenario & 0.35 \\
Upper penetration scenario & 0.45 \\
\hline
\end{tabular}
\end{table}

\vspace{-5pt}
\begin{table}[H]
\captionsetup{font={small}}
\footnotesize
\renewcommand\arraystretch{1}
    \caption{Simulation parameters for the freeway corridor}
    \vspace{-5pt}
    \centering
    \begin{tabular}{llll}
     \hline
         Parameters& Value &Parameters & Value\\ 
          \hline
         $\rho_{j,i}$ &360 (veh/km)&$h_{HC}$& 1.2-1.8 (s)\\\
           $w$& 18 (km/h)& $h_{HH}$& 1.8 (s)\\
          $u_i^{\max}$  & 120 (km/h)& $h_{CC}$& 0.3-0.45 (s)\\
          $u_i^{\min}$  & 40 (km/h)&$h_{CH}$& 1.2-1.8 (s) \\
          $\Delta u_i^{\max}$& 20 (km/h)&$\eta_i$  & 0.2\\
          \hline
    \end{tabular}
    \label{freeway parameters}
\end{table}

\vspace{-10pt}

\begin{figure}[H]
\captionsetup{font={small}}
\centering  
\subfigure[Network average outflow]{
\includegraphics[width=2.5in]{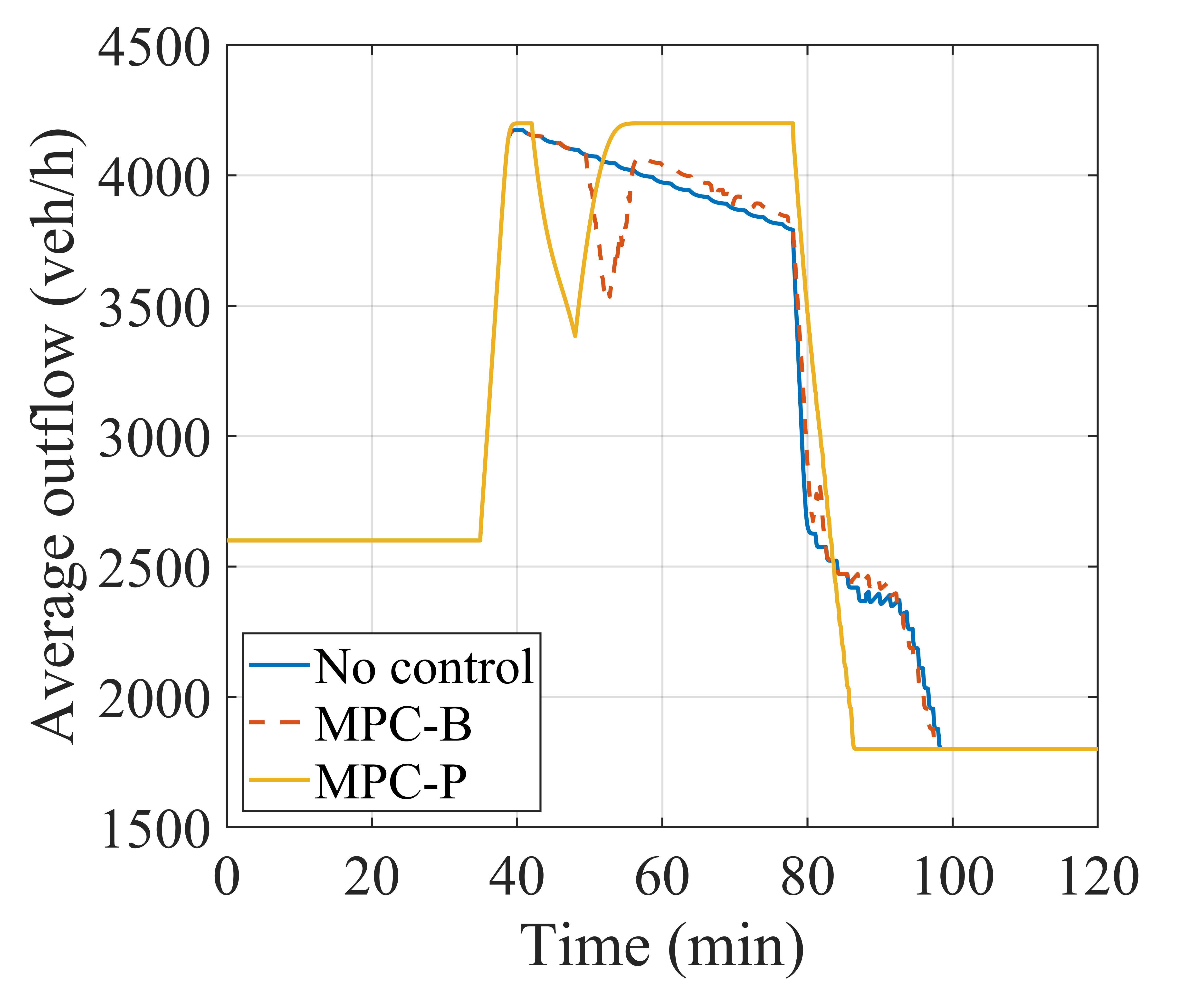}}~~~\subfigure[Total vehicle accumulation]{
\includegraphics[width=2.5in]{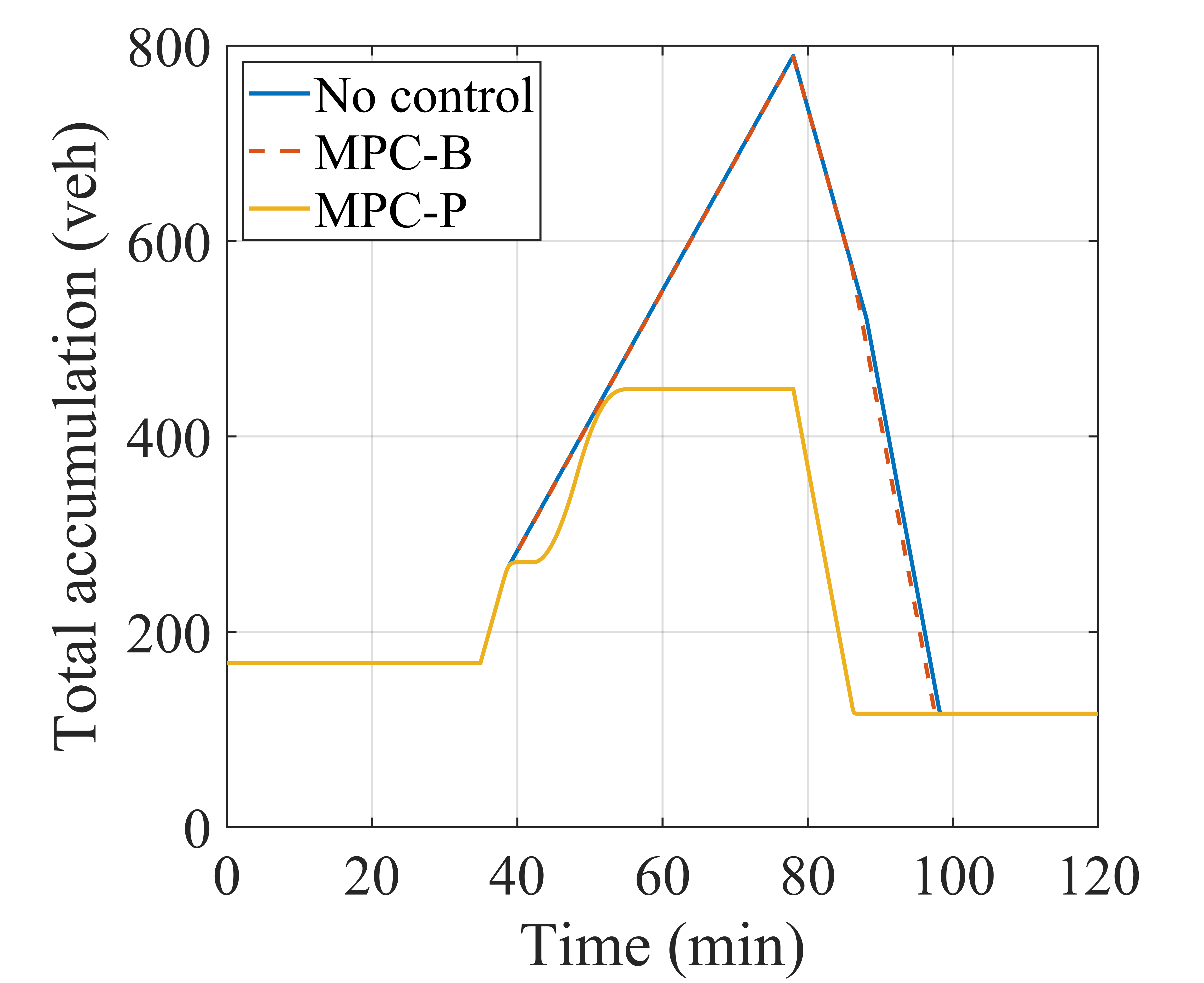}}
\subfigure[Network mean speed]{
\includegraphics[width=2.5in]{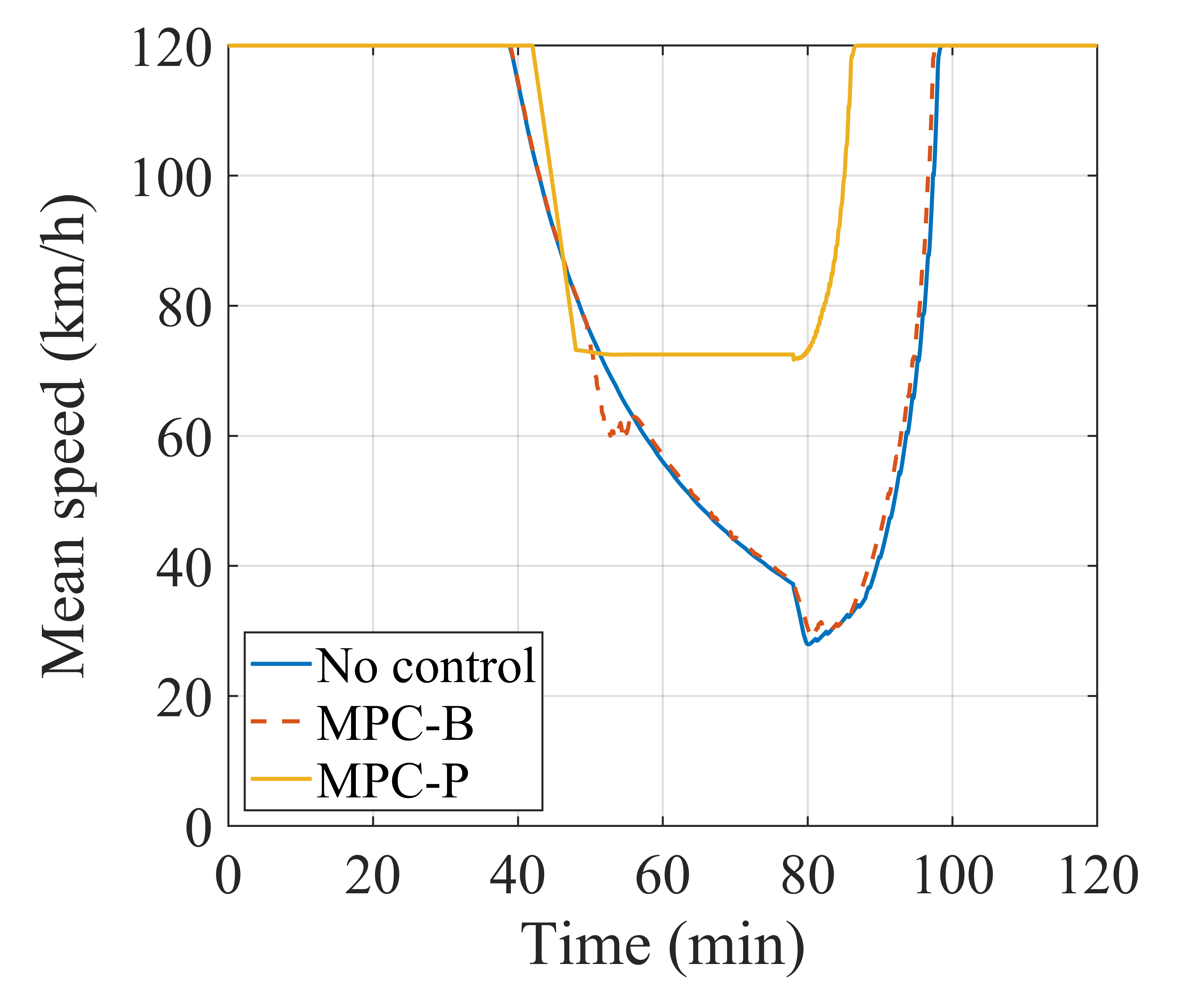}}~~~\subfigure[ Cumulative total time spent]{
\includegraphics[width=2.5in]{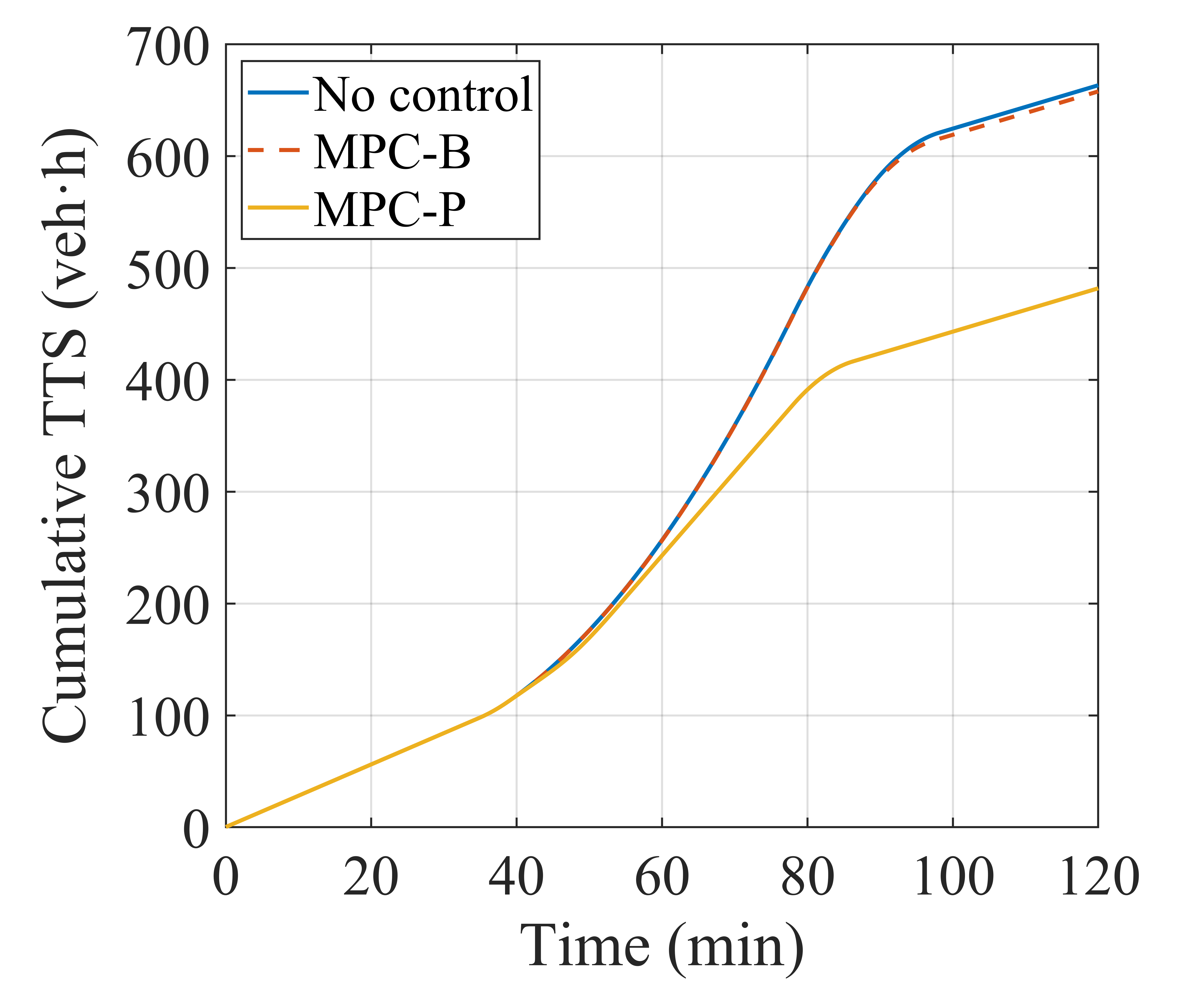}}
\caption{Comparison of network-level operational performance under different freeway control strategies}
\label{c1tts}
\end{figure}

Figure~\ref{c1density} compares the spatiotemporal speed and flow distributions under no control, MPC-B, and MPC-P. Under no control, congestion forms near the downstream bottleneck once the demand exceeds the discharge capacity. The low-speed region then propagates upstream, indicating the formation of a backward-moving congestion wave. The corresponding flow distribution shows a clear reduction in bottleneck discharge after congestion develops, which confirms the occurrence of capacity drop. By dynamically regulating CAV speed limits upstream of the bottleneck, MPC-B alleviates congestion to some extent, leading to a smaller severe congested region. Nevertheless, localized low-speed areas and reduced bottleneck discharge are still observed. In contrast, the proposed MPC-P almost completely eliminates the congestion wave and maintains a higher bottleneck discharge throughout the peak period, demonstrating its superior capability in preventing congestion formation.

The improved performance of MPC-P originates from its explicit consideration of traffic composition uncertainty during traffic prediction and control optimization. In mixed traffic, the penetration rate directly determines the proportion of vehicles that can respond to CAV speed limits. A lower penetration rate reduces the effective control authority while simultaneously altering the mixed traffic fundamental diagram, including the mixed capacity and the condition for capacity drop activation as discussed in Section 
\ref{LTM_Formulation}. Consequently, when MPC-B predicts traffic evolution using only the nominal penetration rate, it may overestimate the effectiveness of speed control and underestimate the likelihood of capacity drop if the actual penetration rate is lower than expected. This mismatch delays the implementation of restrictive speed limits, allowing excessive vehicle accumulation near the bottleneck and eventually triggering congestion. In contrast, MPC-P explicitly accounts for multiple penetration rate scenarios during optimization and identifies a speed limit control strategy that remains effective under unfavorable penetration realizations. Consequently, vehicle accumulation is regulated more proactively, bottleneck discharge remains closer to its maximum level, and the upstream propagation of congestion is effectively suppressed.

Figure~\ref{c1tts} further compares the network-level performance of the three strategies. As shown in Figure~\ref{c1tts}(a), MPC-P maintains the highest network average outflow throughout the peak period, whereas the no control strategy exhibits a significant reduction in outflow due to severe congestion. MPC-B slightly alleviates this degradation but still suffers from significant reduced throughput. Correspondingly, Figure~\ref{c1tts}(b) shows that MPC-P achieves the lowest vehicle accumulation, indicating that it effectively prevents excessive traffic from building up near the bottleneck. Similar trends are observed in the network mean speed and cumulative total time spent (TTS). The MPC-P prevents excessive vehicle accumulation before capacity drop is triggered, resulting in higher average speeds and lower network delay. Note that the MPC-P maintains substantially higher average speeds during the congestion period and enables a faster recovery to free-flow conditions than MPC-B. Consequently, it achieves the lowest cumulative TTS in Figure~\ref{c1tts}(d), confirming that explicitly accounting for traffic composition uncertainty leads to more reliable speed limit control decisions.

\begin{figure}[H]
\captionsetup{font={small}}
\centering  
\subfigure[MPC-B]{
\includegraphics[width=2.55in]{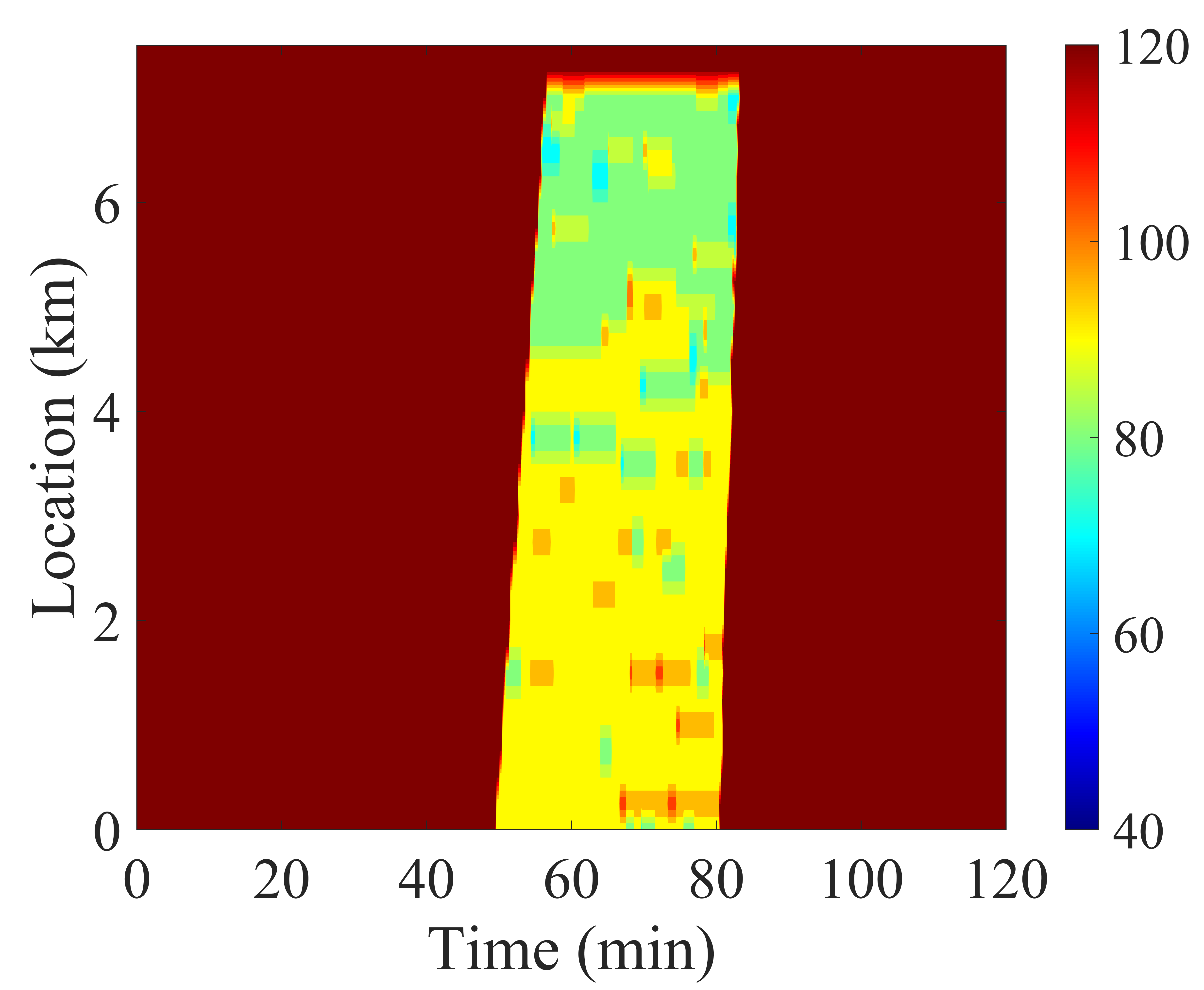}}~~~\subfigure[MPC-P]{
\includegraphics[width=2.55in]{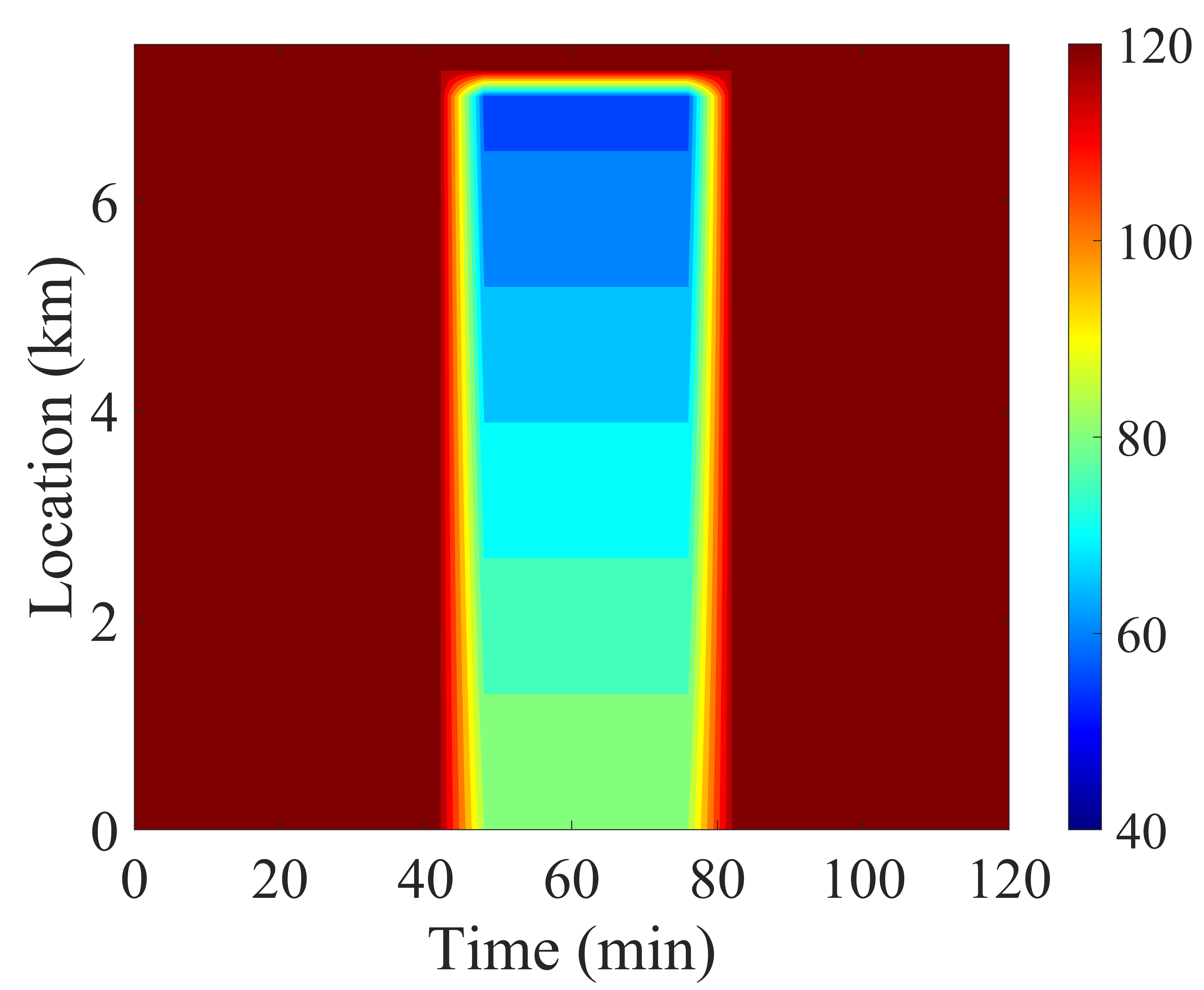}}
\caption{Comparison of spatiotemporal distributions of the optimized CAV speed limits (km/h) along the freeway corridor generated by MPC-B and MPC-P}
\label{c1vsl}
\end{figure}

Figure~\ref{c1vsl} shows the optimized CAV speed limits generated by MPC-B and MPC-P. MPC-B produces speed limits based on the nominal penetration rate and therefore tends to respond after congestion has started to develop. Its control pattern exhibits stronger temporal and spatial fluctuations, indicating that the controller repeatedly adjusts the speed limits as the predicted and realized traffic states diverge. In contrast, MPC-P generates a smoother and more spatially coordinated speed profile. The speed limits gradually decrease toward the downstream bottleneck, forming a controlled speed harmonization zone. This behavior is consistent with the proposed control optimization framework. Rather than optimizing speed limits for a single nominal penetration estimate, MPC-P seeks control actions that remain effective across a range of different penetration rate realizations. Consequently, the optimized speed limits are less sensitive to penetration uncertainty, thereby maintaining stable traffic operation. Moreover, the gradual transitions in speed limits across neighboring links and successive control intervals by the proposed MPC-P improve the practicality of field implementation and enhancing driving consistency.

\subsection{Multi-bottleneck freeway network with merge-diverge spillback}

\begin{figure}
\captionsetup{font={small}}
\centering
\includegraphics[width=6in]{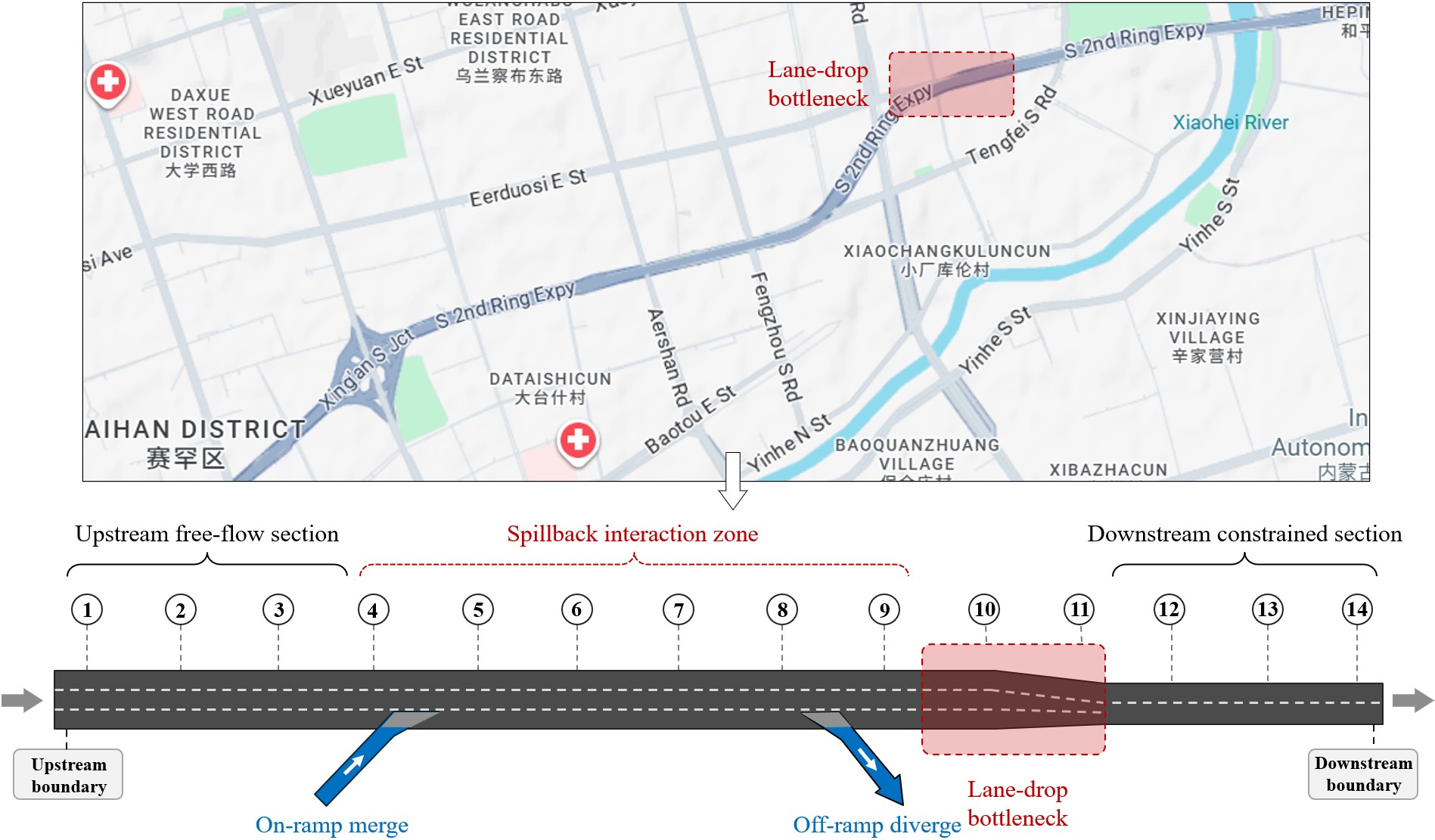}
\caption{Illustration of the Second Ring Expressway in Hohhot, China}
\label{case2}
\end{figure}

\begin{figure}[H]
\captionsetup{font={small}}
\centering
\includegraphics[width=5in]{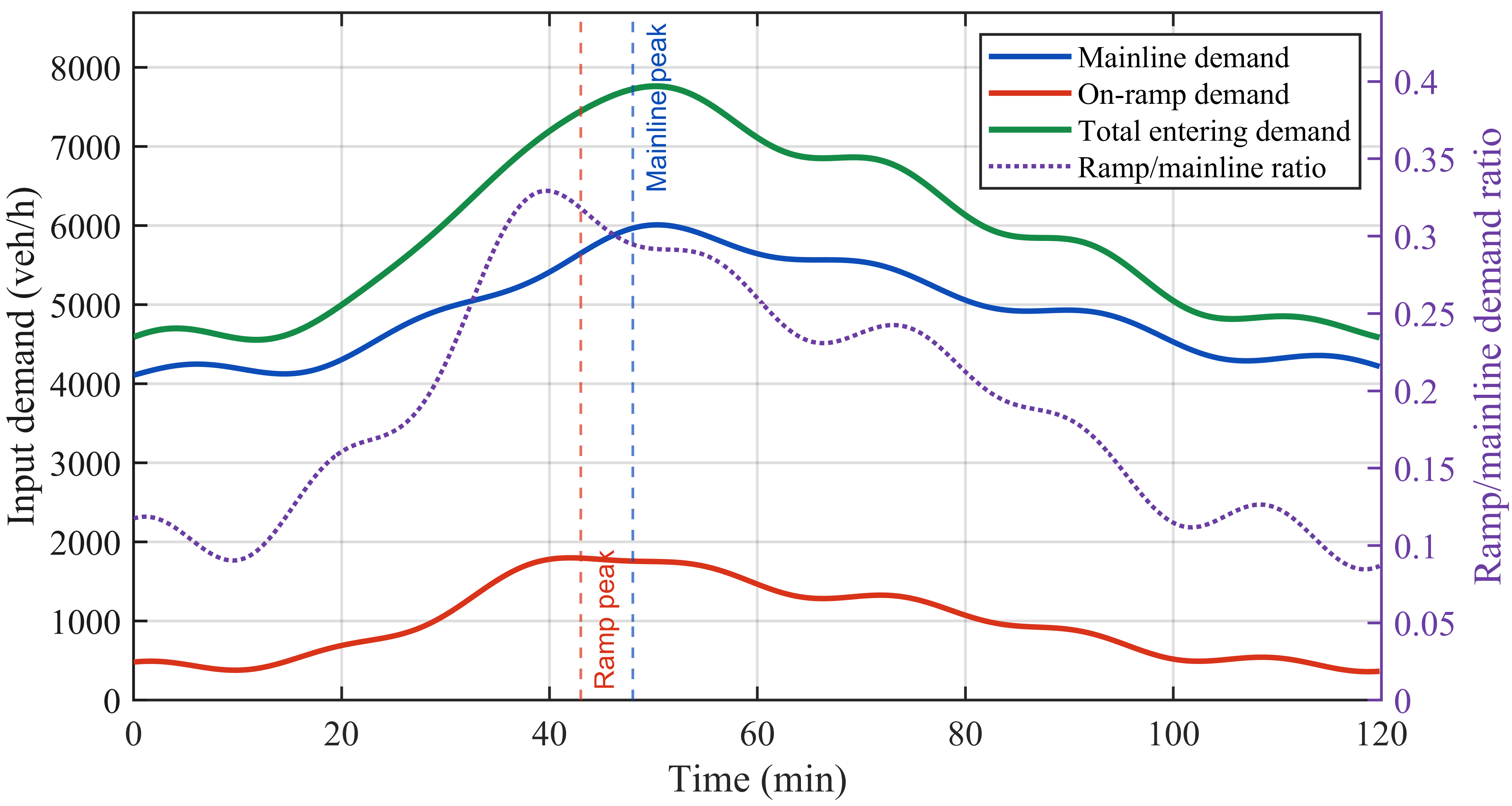}
\caption{Illustration of the traffic demands in the studied freeway network}
\label{demand_case2}
\end{figure}

The second case study evaluates the proposed dynamic speed limit control framework using a real-world freeway network from the Second Ring Expressway in Hohhot, China, as shown in Figure \ref{case2}. The network is divided into 14 links, each 0.5 km long. Table \ref{freeway net parameters} summarizes the simulation network settings. Unlike the single-bottleneck corridor case, this studied network contains multiple interacting bottlenecks associated with on-ramp merging, off-ramp diverging, and lane-drop sections. The downstream bottleneck may generate backward-propagating congestion, which further interferes with the upstream merge area and leads to ramp queue spillback during peak-demand periods. This setting is used to examine whether the proposed controller can maintain robust performance under spatially heterogeneous and time-varying CAV penetration uncertainty. The traffic demands in the studied freeway network are shown in Figure \ref{demand_case2}. 

Three control strategies are compared: no control, deterministic MPC (MPC-B), and the proposed traffic composition-aware MPC (MPC-P). To enable a more direct comparison, MPC-B and MPC-P employ the same objective function in this case, differing only in the treatment of CAV penetration rates. Specifically, the deterministic MPC-B optimizes CAV speed limits based on the nominal estimated penetration rate, while the proposed MPC-P considers multiple admissible penetration rate scenarios. Therefore, the comparison directly quantifies the benefit of incorporating traffic composition uncertainty into the control framework. 

\begin{figure}[H]
\captionsetup{font={small}}
\centering
\includegraphics[width=6.2in]{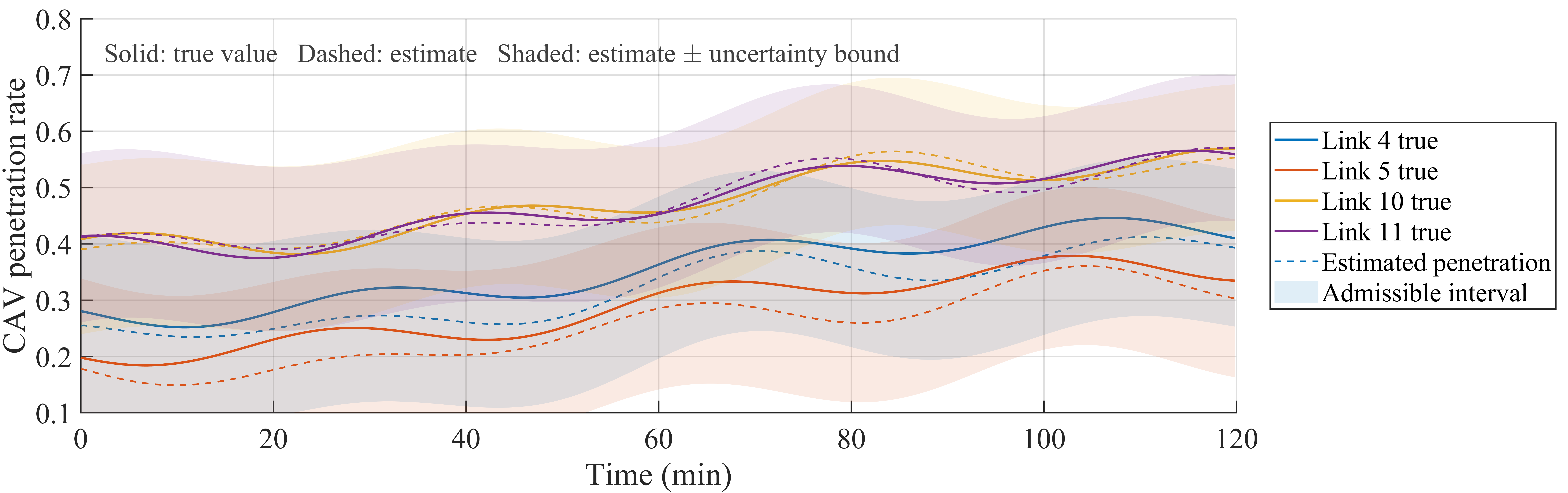}
\caption{CAV penetration rate estimates and uncertainty intervals at critical links in the studied freeway network}
\label{pen bounds}
\end{figure}

Figure~\ref{pen bounds} illustrates the time-varying CAV penetration rates and the corresponding uncertainty bounds in this case. The solid curves represent the true penetration rates at selected critical links, while the dashed curves denote the estimated penetration rates available to the controller. The shaded regions indicate the admissible uncertainty intervals. It can be observed that the estimated penetration rates deviate from the true values. The deviation is more pronounced around the merge and lane-drop bottleneck regions, where traffic composition changes rapidly. This setting provides a more realistic test environment than assuming a uniform constant penetration rate error across all links and all times. The simulation lasts 120 min. The traffic state is updated every 10 s. 

\begin{table}[H]
\captionsetup{font={small}}
\footnotesize
\renewcommand\arraystretch{1}
    \caption{Simulation parameters for the multi-bottleneck freeway network}
    \vspace{-5pt}
    \centering
    \begin{tabular}{llll}
     \hline
         Parameters& Value &Parameters & Value\\ 
          \hline
         $\rho_{j,i}$ &150 (veh/km/ln)&$h_{HC}$& 1.2-1.8 (s)\\\
           $w$& 18 (km/h)& $h_{HH}$& 1.8 (s)\\
          $u_i^{\max}$  & 120 (km/h)& $h_{CC}$& 0.3-0.45 (s)\\
          $u_i^{\min}$  & 50 (km/h)&$h_{CH}$& 1.2-1.8 (s) \\
          $\Delta u_i^{\max}$& 20 (km/h)&Averaged $\eta_i$  & 0.3\\
          \hline
    \end{tabular}
    \label{freeway net parameters}
\end{table}

\begin{figure}[H]
\captionsetup{font={small}}
\centering
\includegraphics[width=4.9in]{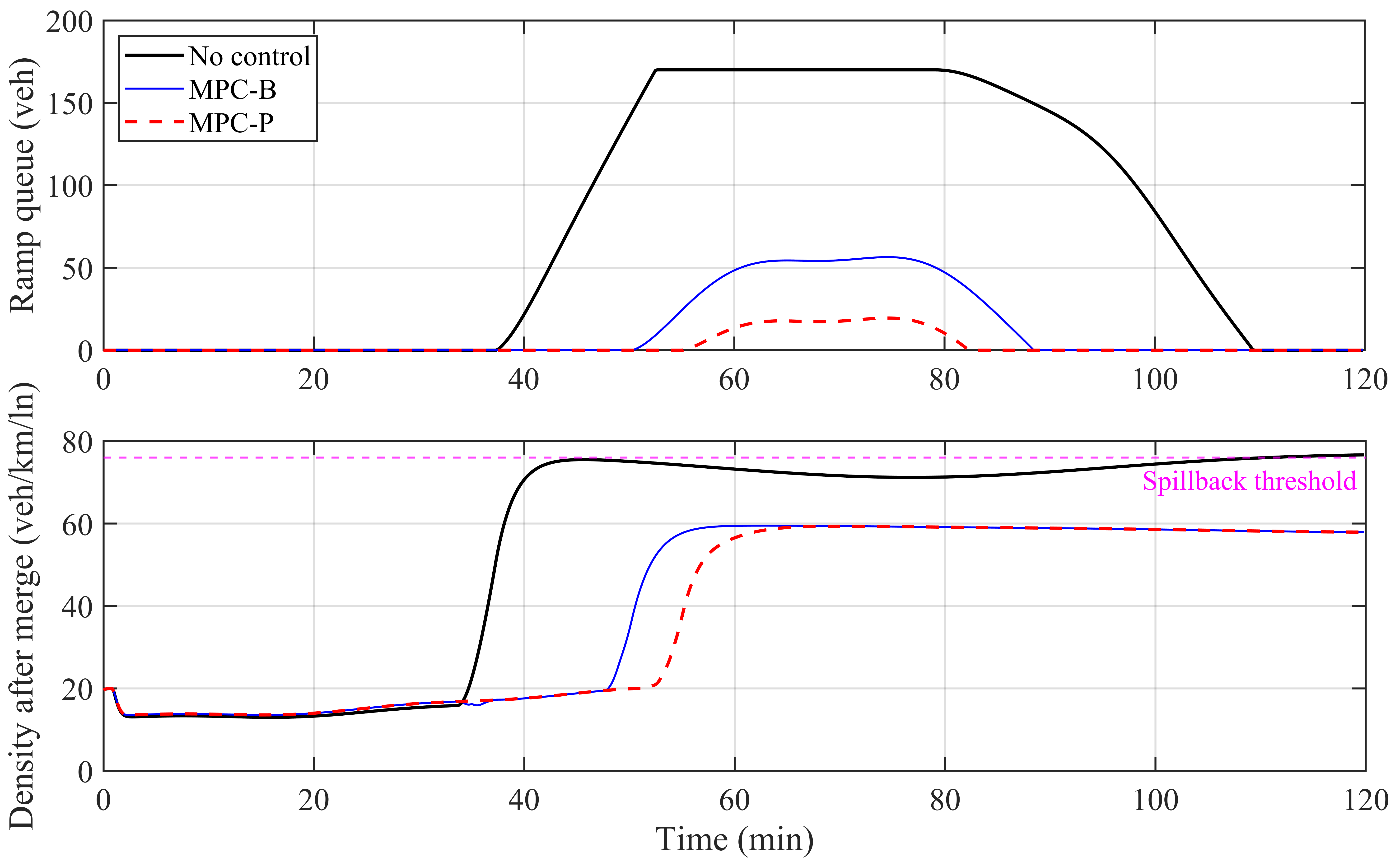}
\caption{Comparison of ramp queue evolution and spillback risk under different control  strategies}
\label{queuecase2}
\end{figure}

\begin{figure}[H]
\captionsetup{font={small}}
\centering  
\subfigure[Capacity drop exposure at the bottleneck]{
\includegraphics[width=2.5in]{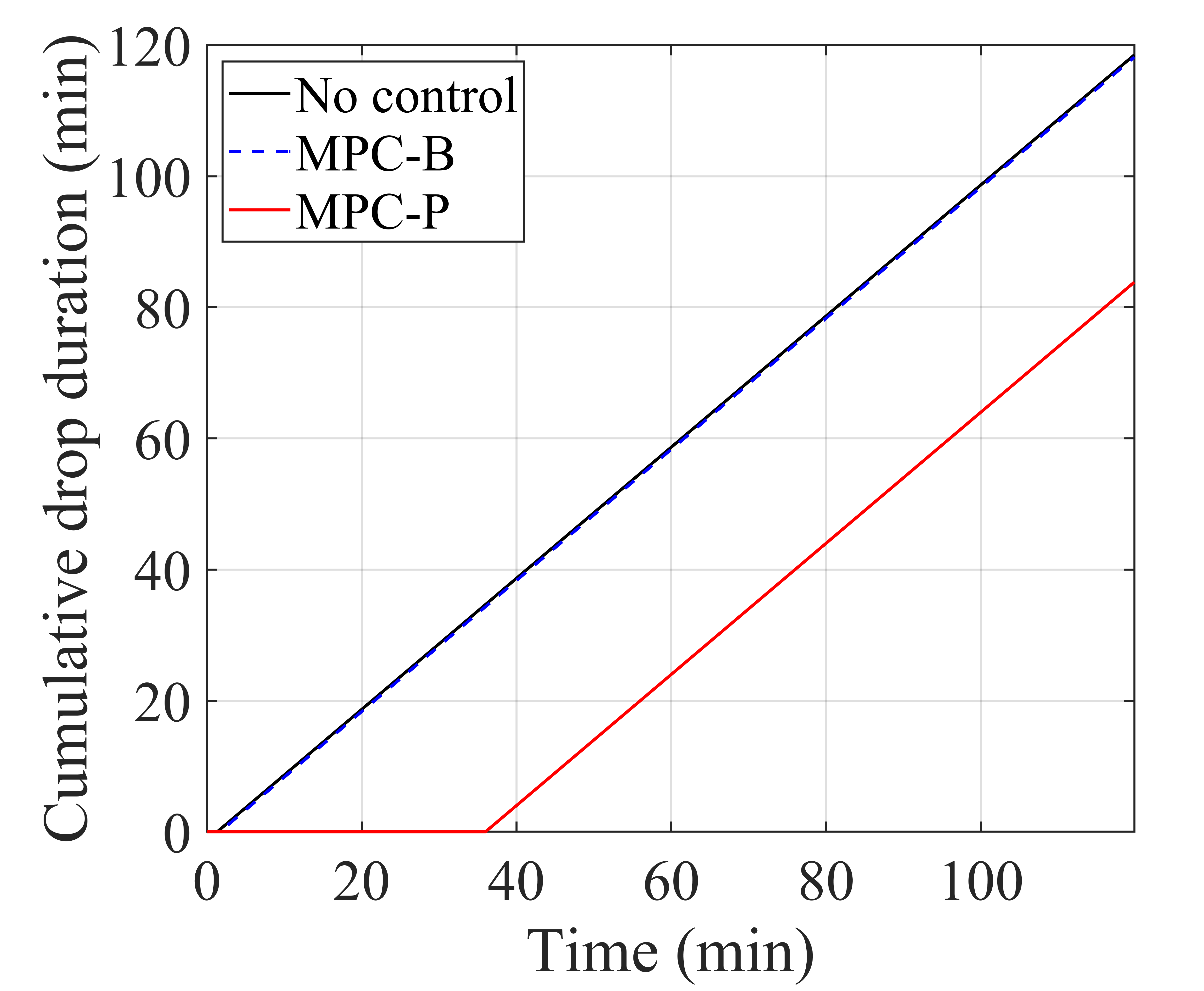}}~~~\subfigure[Improvement over MPC-B]{
\includegraphics[width=2.5in]{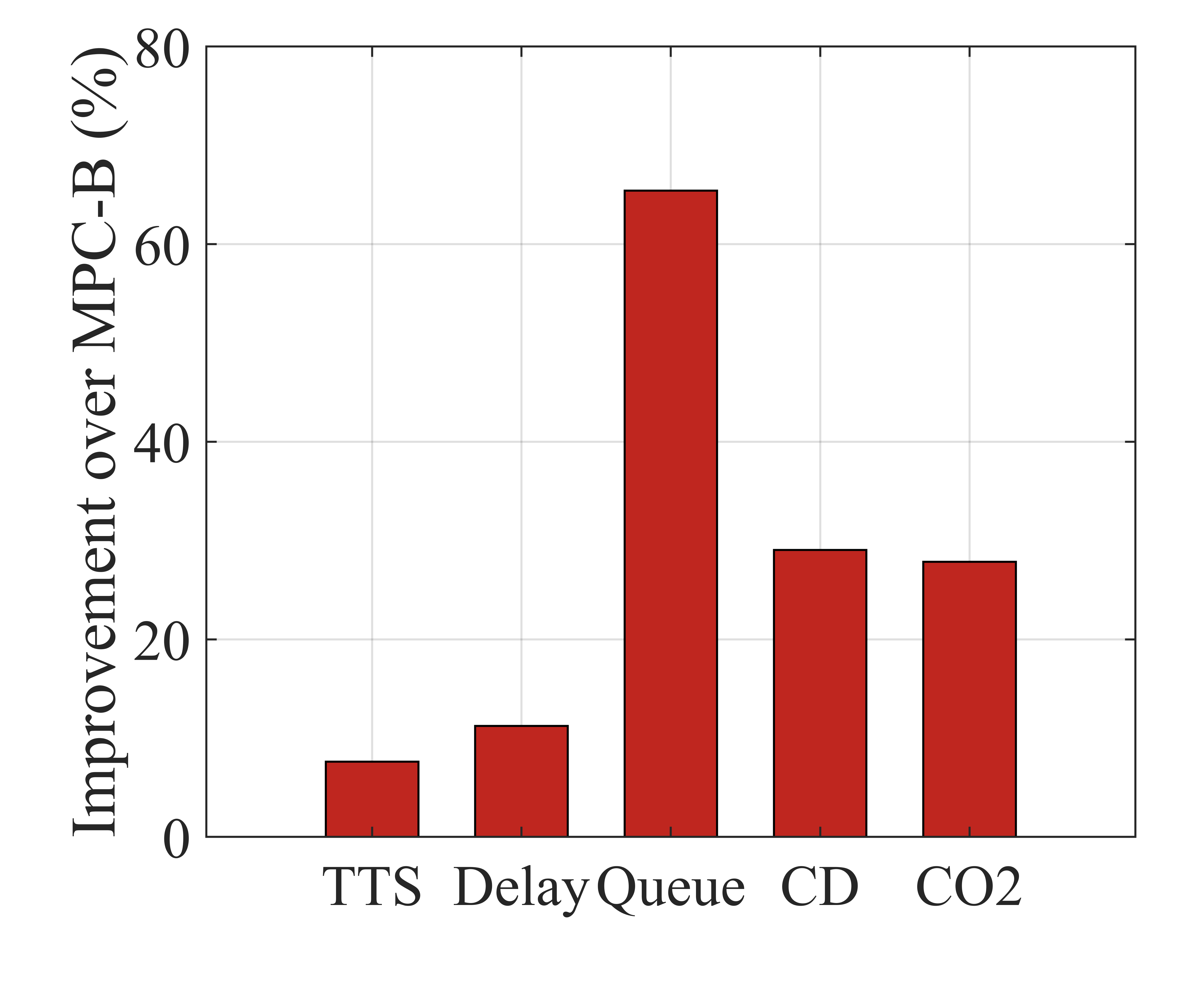}}
\caption{Illustration of the capacity drop (CD) exposure and improvement at the bottleneck}
\label{capacitydropcase2}
\end{figure}

The performance of the proposed approach is evaluated in terms of TTS, travel delay, queue length, capacity drop, and emissions. It should be noted that the delay and emission-related indicators reported in this case study are macroscopic performance proxies derived from the simulated traffic states. The delay is calculated based on network accumulation rather than individual vehicle trajectories. Specifically, the total number of vehicles in the network at time step $k$ is computed as
\begin{equation}
N(k)=\sum_{i=1}^{m}\rho_i(k)L_i n_i + Q_r(k),
\end{equation}
where $\rho_i(k)$ is the density of link $i$, $L_i$ is the link length, $n_i$ is the number of lanes, and $Q_r(k)$ is the on-ramp queue. A free-flow reference accumulation is defined as
\begin{equation}
N_{\mathrm{ff}}=\sum_{i=1}^{m}\rho_{\mathrm{ff}}L_i n_i,
\end{equation}
where $\rho_{\mathrm{ff}}$ denotes the reference free-flow density. The accumulation-based delay is then calculated as
\begin{equation}
D=\sum_{k}\max\left\{0,N(k)-N_{\mathrm{ff}}\right\}\Delta t.
\end{equation}

Therefore, this delay indicator measures the excess vehicle accumulation relative to a free-flow reference state, and should be interpreted as a macroscopic delay proxy. An emission-related proxy is also introduced to compare the environmental implication of different control strategies \citep{zegeye2013integrated}. Since the simulation is conducted at the macroscopic link level, calibrated vehicle-level emission models such as MOVES or VT-Micro are not directly applied. Instead, the proxy is formulated as
\begin{equation}
E(k)=
\sum_{i=1}^{m}
\left(
\alpha_0+\alpha_1 v_i^2(k)+\alpha_2 a_i^2(k)
\right)
\rho_i(k)L_i n_i,
\end{equation}
where $v_i(k)$ is the average speed of link $i$, and $a_i(k)$ is approximated by the spatial speed variation between neighboring links. The total emission-related proxy is given by
\begin{equation}
E_{\mathrm{total}}=\sum_k E(k)\Delta t.
\end{equation}

This indicator is not intended to represent absolute CO$_2$ emissions in physical units. Rather, it provides a consistent relative comparison of traffic accumulation, speed level, and speed disturbance across different control strategies. The coefficients $\alpha_0,\alpha_,\alpha_2$ are normalized weighting parameters used to balance the relative contributions of traffic accumulation, speed level, and speed disturbance.

Figure~\ref{queuecase2} first compares the ramp queue and the density after the merge. Under no control, the ramp queue reaches the storage limit of 170 vehicles and spillback lasts for 44.17 min. Both MPC strategies eliminate spillback, demonstrating the effectiveness of CAV speed limit control in regulating mainline inflow and protecting the merge area. However, the proposed MPC-P provides a much stronger queue reduction than the MPC-B. The maximum ramp queue decreases from 56.44 vehicles under MPC-B to 19.51 vehicles under MPC-P, corresponding to a reduction of approximately 65.4\%. This indicates that the proposed MPC-P not only avoids full spillback, but also maintains a larger operational buffer at the merge area. The cumulative capacity drop exposure further confirms this improvement as shown in Figure \ref{capacitydropcase2}(a). The MPC-B still experiences 118.17 min of capacity drop exposure at the critical bottleneck, which is close to the no control case. In contrast, the proposed MPC-P reduces the capacity drop duration to 83.83 min, a 29.1\% reduction compared with the MPC-B. This shows that the proposed MPC-P does not merely shift queues upstream, but actively reduces the duration of bottleneck breakdown. 

\begin{figure}
\captionsetup{font={small}}
\centering
\includegraphics[width=5.4in]{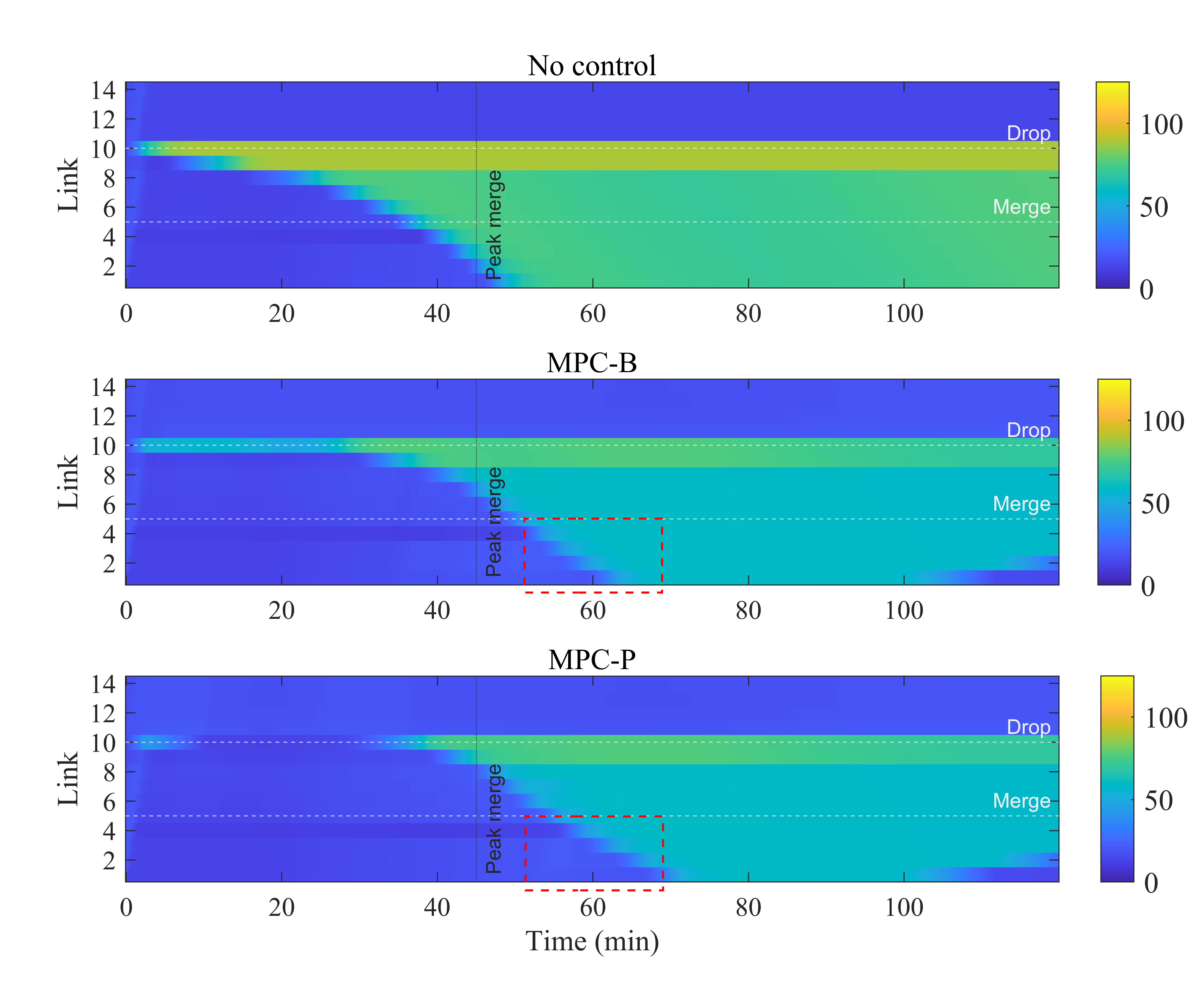}
\caption{Comparison of spatiotemporal density (veh/km) distributions under  different control strategies}
\label{dencase2}
\end{figure}

\begin{figure}
\captionsetup{font={small}}
\centering
\includegraphics[width=5.4in]{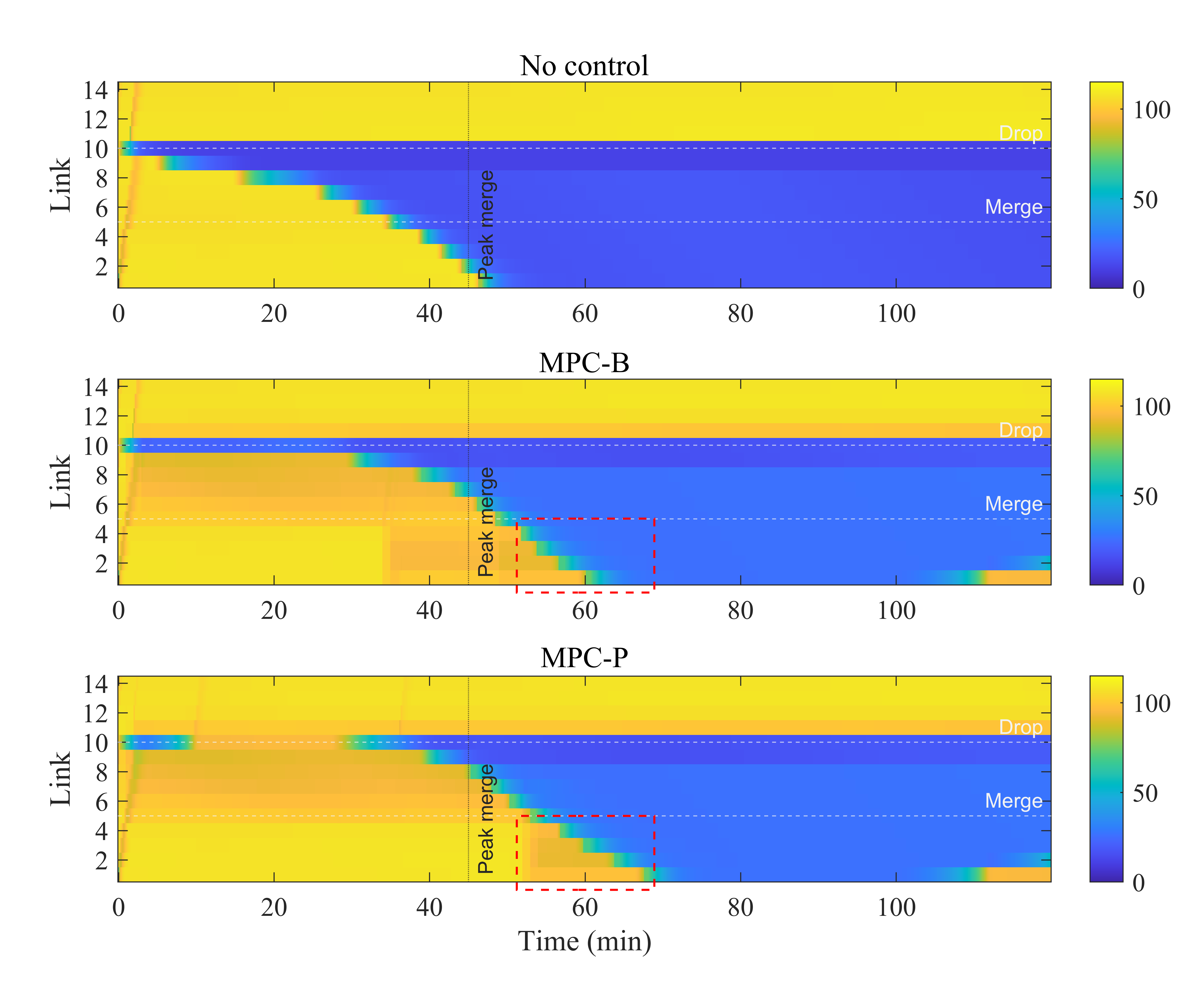}
\caption{Comparison of spatiotemporal speed (km/h) distributions under different control strategies}
\label{speedcase2}
\end{figure}

Figures \ref{dencase2} and \ref{speedcase2} show the spatiotemporal density and speed distributions of the freeway mainline under the three strategies. Under the no control scenario, congestion forms near the downstream lane-drop bottleneck and propagates upstream. As the queue reaches the merge area, the on-ramp discharge is restricted and a severe ramp queue is generated as presented in Figure \ref{queuecase2}. This confirms the existence of strong interaction between the downstream bottleneck and the upstream merge bottleneck. The MPC-B mitigates congestion propagation by applying CAV speed regulation upstream of the bottleneck. Compared with no control, the high-density region is substantially reduced and spillback is avoided. However, because MPC-B relies on a nominal penetration rate estimate, its predicted traffic evolution may deviate from the actual mixed condition. As a result, capacity drop exposure remains long  as shown in Figure \ref{capacitydropcase2}(a), and non-negligible ramp queues still appear during peak period.




\begin{figure}
\captionsetup{font={small}}
\centering  
\subfigure[Density difference]{
\includegraphics[width=2.6in]{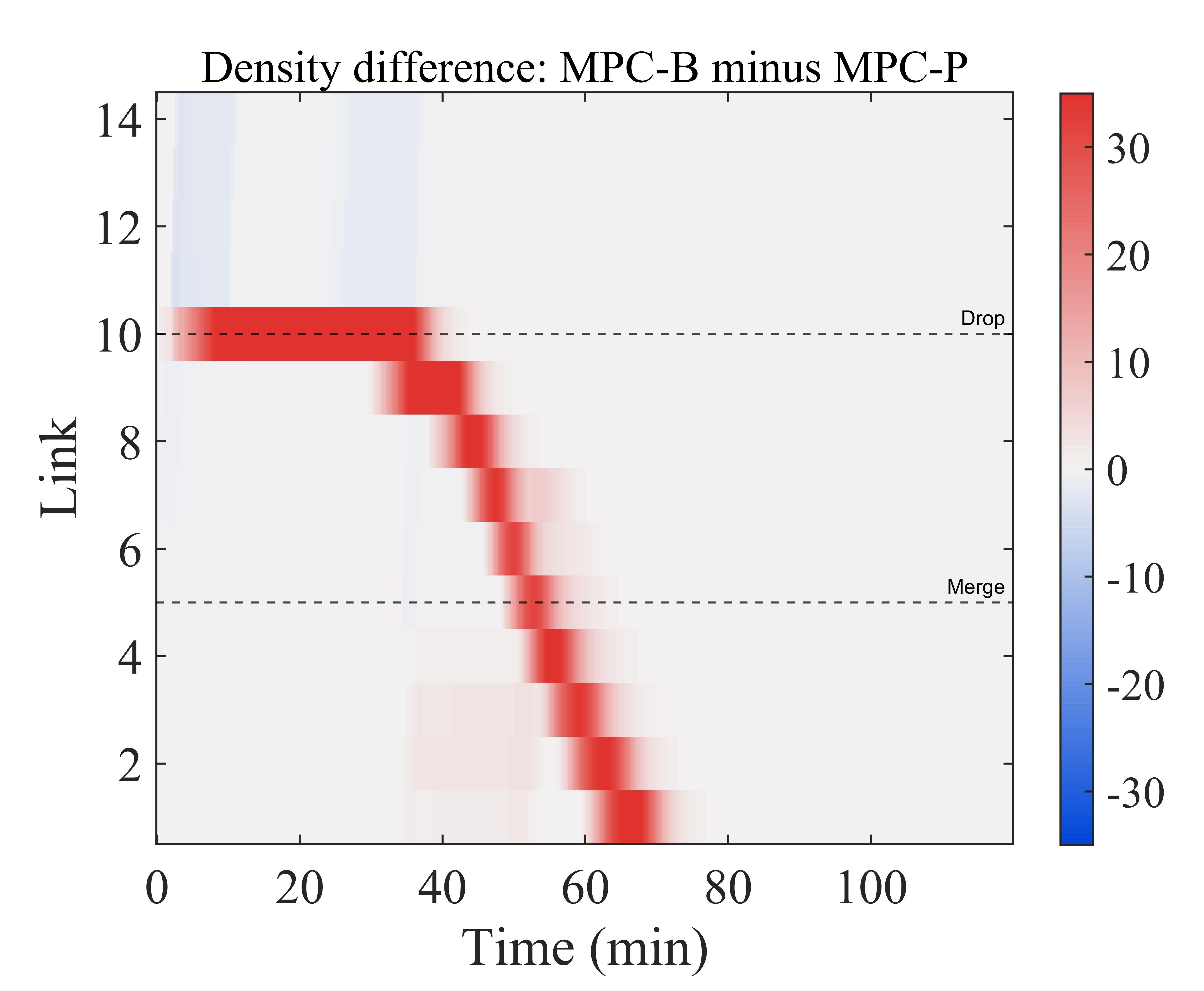}}~~\subfigure[Speed difference]{
\includegraphics[width=2.6in]{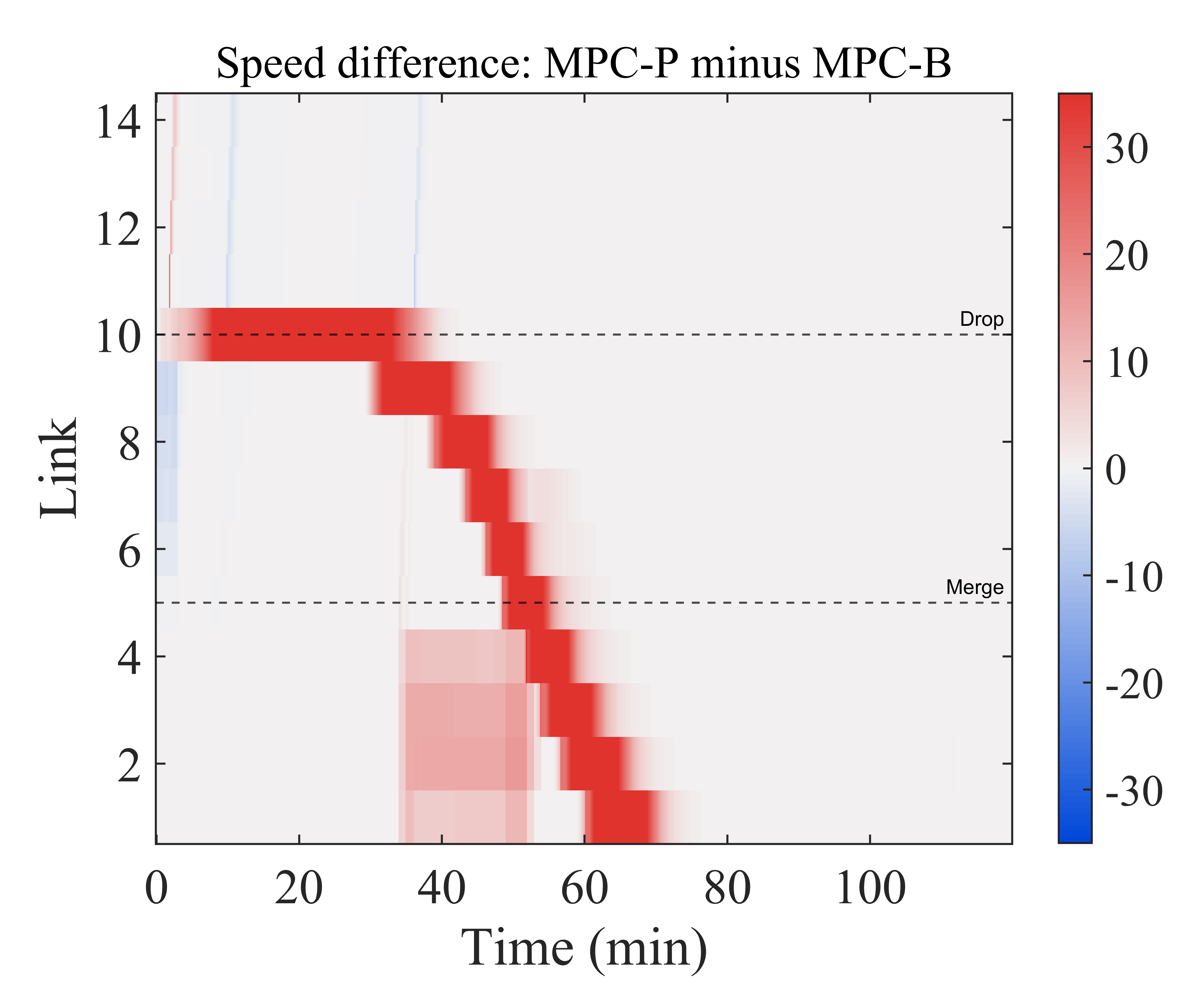}}
\vspace{-3pt}
\caption{Comparison of density and speed differences between MPC-B and MPC-P}
\label{denccase2}
\end{figure}
\begin{figure}
\captionsetup{font={small}}
\centering  
\subfigure[Cumulative total time spent]{
\includegraphics[width=2.66in]{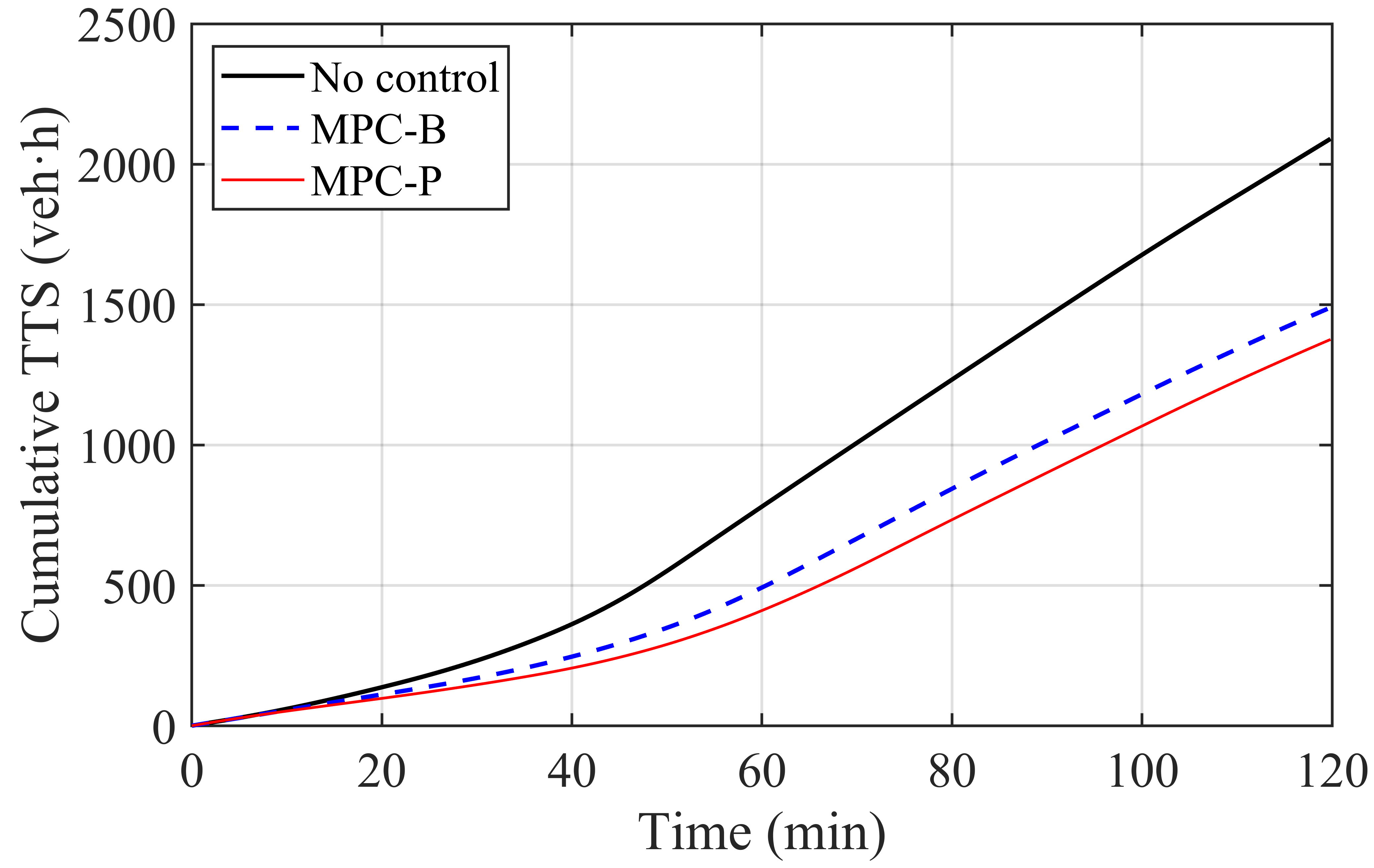}}~~~\subfigure[Cumulative delay]{
\includegraphics[width=2.66in]{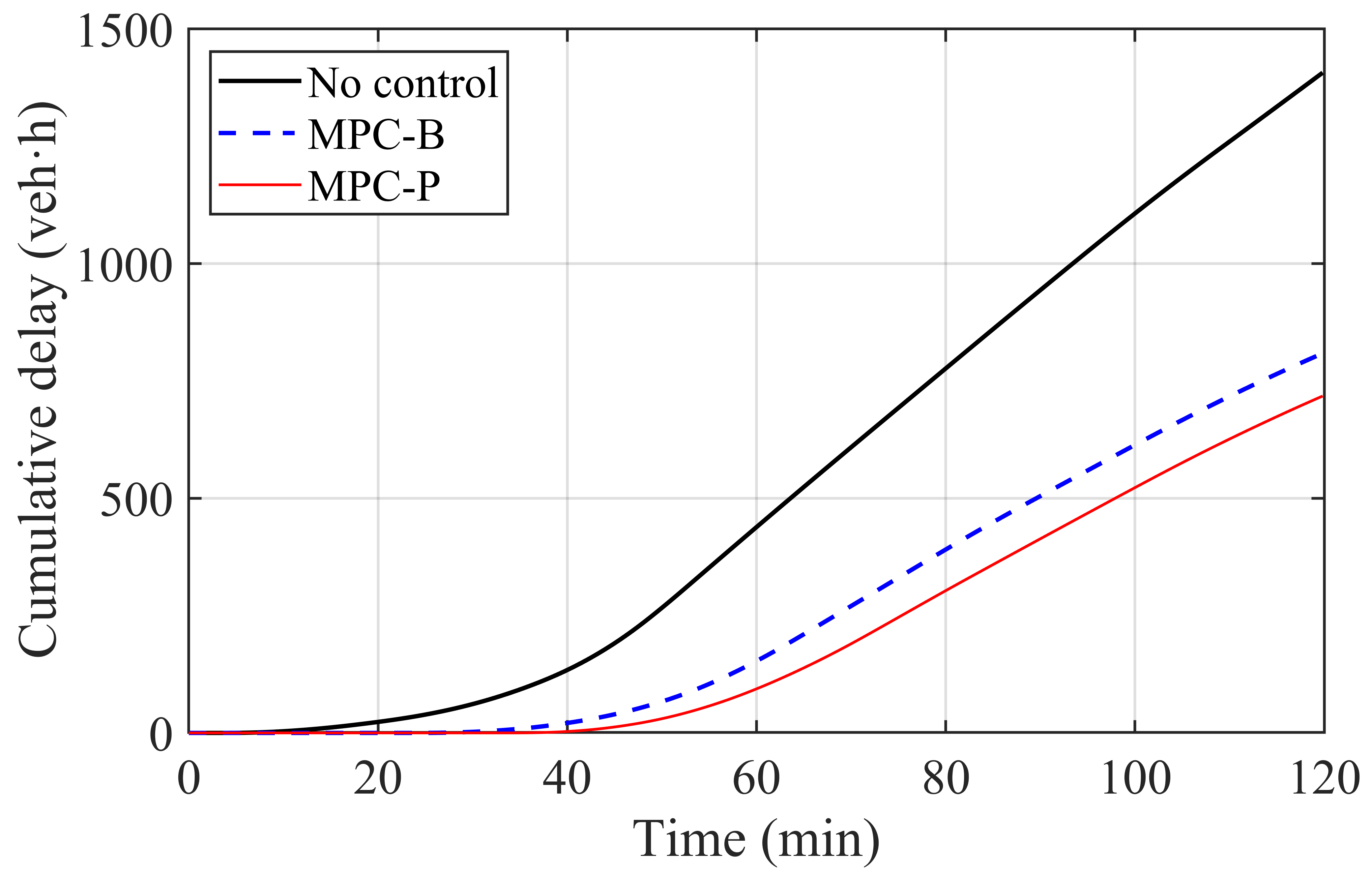}}
\vspace{-3pt}
\caption{Comparison of the cumulative total time spent and delay by different control strategies}
\label{cumcase2}
\end{figure}
\begin{figure}
\captionsetup{font={small}}
\centering
\includegraphics[width=5.6in]{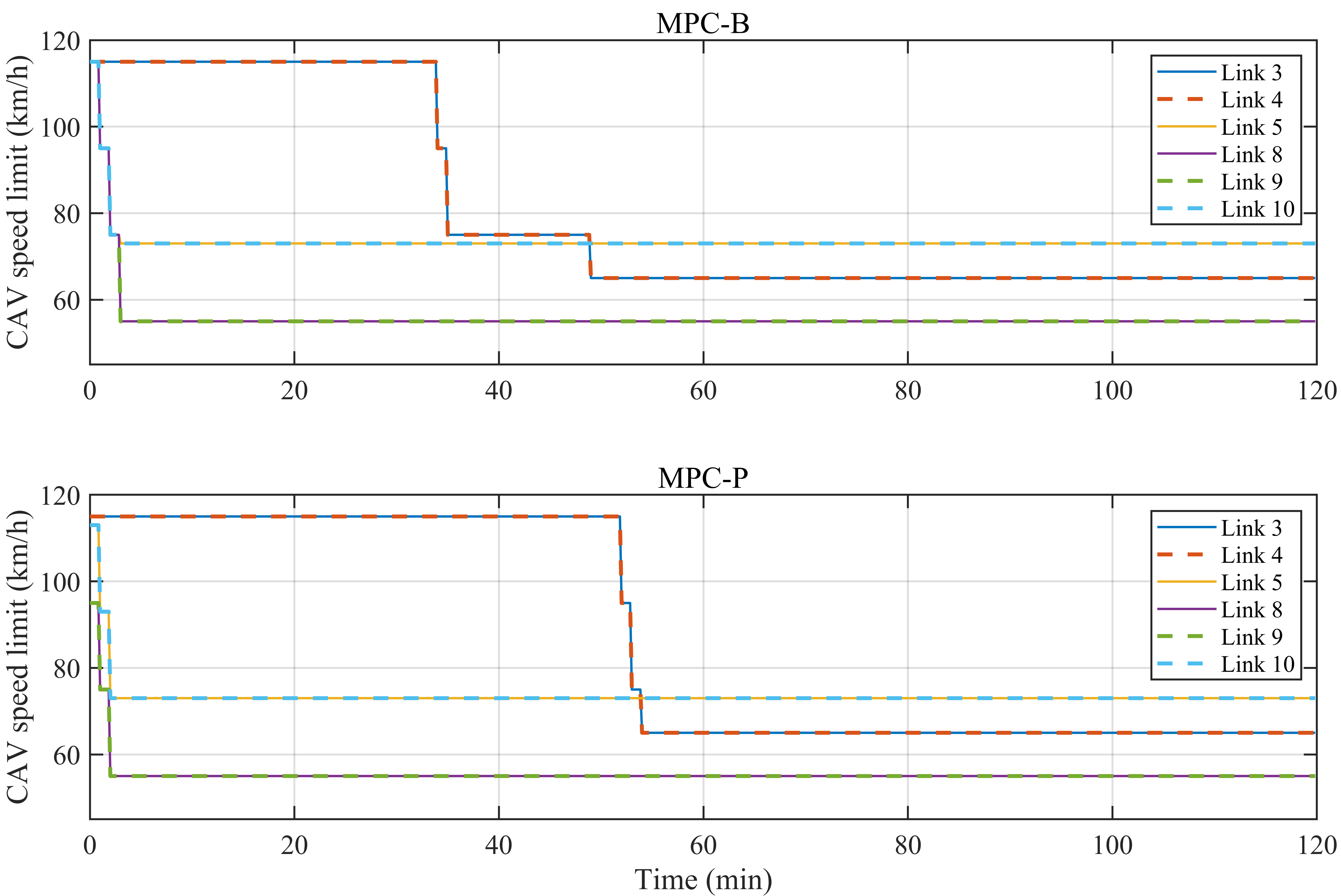}
\caption{Comparison of the optimized CAV speed limits at critical links generated by MPC-B and
MPC-P}
\label{VSLcase2}
\end{figure}

Figures \ref{denccase2} and \ref{cumcase2} further compare the differences of density and speed  between MPC-B and MPC-P. The results show that MPC-P reduces vehicle accumulation mainly around the lane-drop bottleneck and along the backward-propagating congestion front. Meanwhile, the speed difference plot shows that the proposed MPC-P maintains higher speeds in the same critical regions, indicating that the proposed MPC-P prevents the traffic state from entering a severe congested regime. This result highlights the benefits of involving traffic composition uncertainty in the MPC framework. The proposed MPC-P considers unfavorable penetration scenarios and therefore applies speed harmonization earlier as shown in Figure \ref{VSLcase2}. This allows the controller to reduce the risk of bottleneck breakdown and maintain more stable discharge conditions.

Table~\ref{tab:caseB_summary} summarizes the system-level performance. The no control strategy results in the worst performance, with a total TTS of 2090.48 veh$\cdot$h and a total delay of 1406.95 veh$\cdot$h. Deterministic MPC-B reduces TTS to 1489.67 veh$\cdot$h and delay to 808.86 veh$\cdot$h by preventing severe spillback. The proposed MPC-P achieves the best overall performance. Compared with the MPC-B, the proposed MPC-P reduces TTS from 1489.67 to 1376.04 veh$\cdot$h, corresponding to a 7.6\% improvement. Total delay is reduced from 808.86 to 717.86 veh$\cdot$h, corresponding to an 11.2\% improvement. The CO$_2$ proxy is also reduced by 27.8\%, suggesting that the smoother and more stable traffic evolution generated by the MPC-P can provide environmental benefits.

\begin{table}[H]
\captionsetup{font={small}}
\footnotesize
\centering
\caption{Overview of the performance comparison in the freeway network with merge-diverge spillback}
\label{tab:caseB_summary}
\begin{tabular}{lccccc}
\hline
Strategy & TTS & Delay & Max queue & Spillback & Capacity drop \\
 & (veh$\cdot$h) & (veh$\cdot$h) & (veh) & (min) & (min) \\
\hline
No control & 2090.48 & 1406.95 & 170.00 & 44.17 & 118.50 \\
MPC-B & 1489.67 & 808.86 & 56.44 & 0.00 & 118.17 \\
MPC-P & 1376.04 & 717.86 & 19.51 & 0.00 & 83.83 \\
\hline
\end{tabular}
\end{table}

\begin{figure}[H]
\captionsetup{font={small}}
\centering
\includegraphics[width=6.3in]{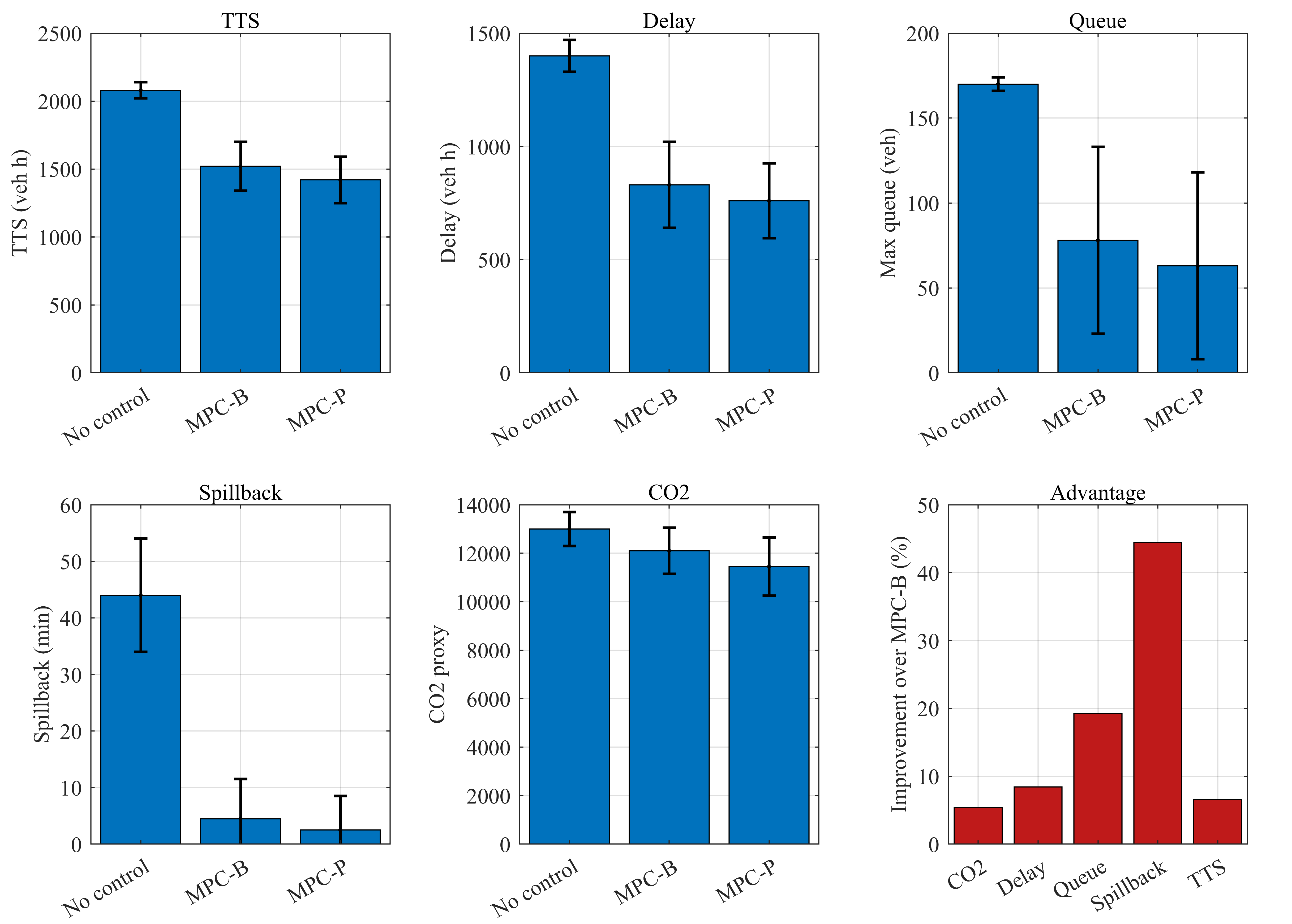}
\caption{Robustness test of the proposed MPC-P in the studied freeway network}
\label{robustnes}
\end{figure}

To further examine the robustness of the proposed MPC-P, a Monte Carlo experiment was conducted by perturbing demand intensity, demand peak timing,  and bottleneck capacity. For each random realization, the no control, the MPC-B, and the proposed MPC-P strategies were simulated under the same traffic condition. As shown in Figure \ref{robustnes}, the blue bars represent the sample mean over 24 random realizations, and the error bars indicate standard deviation. Although the improvement in the Monte Carlo average is smaller than that in the baseline simulation without random perturbations, the proposed MPC-P still provides more reliable performance. The propsoed MPC-P achieves lower average TTS, delay, maximum ramp queue, spillback duration, and emission-related proxy than the MPC-B. Specifically, the mean TTS decreases from 1520 veh$\cdot$h under MPC-B to 1422 veh$\cdot$h under MPC-P. The mean delay decreases from 832 to 760 veh$\cdot$h, and the mean maximum ramp queue decreases from 78 to 63 vehicles. The proposed MPC-P also reduces the mean CO$_2$ proxy from abouot 12100 to 11430. This result is expected because robust control is designed in MPC-P to protect the system from unfavorable uncertainty realizations rather than optimize only the nominal case. Consequently, the proposed MPC framework is particularly valuable when the estimated CAV penetration rate is uncertain and the freeway network contains interacting bottlenecks.

Therefore, the results of this case study demonstrate that the main advantage of the proposed MPC-P is not simply reducing density everywhere in the freeway network. Instead, its benefit lies in controlling where and when congestion is allowed to form. By considering traffic composition uncertainty, the robust controller prevents excessive accumulation near the merge and lane-drop bottlenecks, reduces capacity drop exposure, and avoids spillback. This behavior is important for freeway networks with interacting bottlenecks, where a local prediction error may trigger network-level congestion propagation. Overall, this case study confirms that the proposed MPC-P framework can provide system-level operational benefits under uncertain mixed traffic conditions. 

The computational efficiency of the proposed MPC-P was also evaluated. The wall-clock solution time was recorded at each rolling control update. The control interval was 60 s, and both MPC-B and MPC-P were updated 120 times over the simulation horizon. Since no optimization problem is solved in the no control case, its computation time is zero. As shown in Table~\ref{tab:case2_computation_time}, the MPC-B required 0.0197 s on average per control update, with a maximum solution time of 0.0401 s. The proposed MPC-P required 0.0533 s on average, with a maximum solution time of 0.0697 s. Although the MPC-P evaluates multiple penetration rate scenarios and therefore introduces additional computational effort, its average solution time remains much smaller than the 60 s control interval. The corresponding real-time ratio is only 0.00089, indicating that the proposed scenario-based MPC-P is computationally feasible for efficient speed limit control in the studied multi-bottleneck network. 

Note that although the computational cost $\mathcal{O}$ of the proposed MPC-P scales approximately linearly with the number of penetration rate scenarios $\mathcal{O}
\propto
N_s$. Since the prediction model is macroscopic and link-based, the proposed method still computationally tractable than the microscopic vehicle-based MPC and suitable for rolling-horizon freeway traffic control.

\begin{table}[H]
\captionsetup{font={small}}
\footnotesize
\centering
\caption{Computational time of the controllers in the studied freeway network}
\label{tab:case2_computation_time}
\begin{tabular}{lccccc}
\hline
Control strategy & Updates & Total time (s) & Mean time (s) & Max time (s) & Real-time ratio \\
\hline
No control & 0 & 0 & 0 & 0 & 0 \\
MPC-B & 120 & 2.3665 & 0.0197 & 0.0401 & 0.00033 \\
MPC-P & 120 & 6.3952 & 0.0533 & 0.0697 & 0.00089 \\
\hline
\end{tabular}
\end{table}

\section{Conclusion}
\label{Conclusion}
This study proposed a link-based MPC framework for dynamic CAV speed limit control in freeway networks with mixed traffic flow. Unlike existing control approaches that assume deterministic mixed traffic characteristics, this work explicitly considered the traffic composition uncertainty associated with the time-varying CAV penetration rate, and incorporated it into both traffic flow prediction and control optimization. A traffic composition-aware MPC framework was developed to generate CAV speed limits that remain effective under multiple admissible penetration rate realizations. The simulation results from both the freeway corridor and the real-world freeway network demonstrate that explicitly accounting for traffic composition uncertainty leads to more reliable flow prediction and more effective speed control decisions. Rather than reacting after congestion has developed, the proposed controller proactively regulates vehicle accumulation upstream of bottlenecks, thereby reducing the likelihood of capacity drop and suppressing congestion propagation. Compared with the existing deterministic MPC, the proposed approach consistently produces smoother and more spatially coordinated speed limits, resulting in improved traffic efficiency and more stable freeway operations under heterogeneous mixed traffic conditions.

The findings of this study suggest that traffic composition uncertainty deserves explicit consideration in model-based mixed traffic control. Explicitly modeling how penetration rate uncertainty propagates through the mixed fundamental diagram provides a physically interpretable approach for improving the robustness of mixed traffic flow prediction and control. As connected vehicle technologies continue to evolve, such uncertainty-aware control strategies are expected to become increasingly important for real-world freeway traffic management.

Future work will focus on integrating online penetration rate estimation into the proposed framework, allowing the uncertainty set to be updated adaptively using real-time connected vehicle observations. Another promising direction is to combine the proposed uncertainty-aware modeling framework with learning-based traffic prediction and coordinated freeway control strategies, including ramp metering and lane management \citep{han2026cooperative,jin2026optimal}. In addition, future field implementation will be necessary to further validate the applicability of the proposed approach in complex traffic environments.

\section*{Acknowledgment}
This research has been supported by the National Natural Science Foundation of China (No. T2588101). 

\bibliographystyle{elsarticle-harv}
\bibliography{Reference}

\vfill

\end{document}